\documentclass[12pt,english]{article}

\usepackage[english]{babel}
\usepackage{amsmath,amsthm}
\usepackage[utf8]{inputenc}
\usepackage{bm}%
\usepackage[colorlinks=true,linkcolor=blue]{hyperref}%
 \usepackage{natbib}
\usepackage{graphicx,verbatim,array,palatino,amssymb, epsfig,multirow} 
\usepackage{color}
\usepackage{psfrag}
\usepackage{textcomp}
\usepackage{tikz}
\usepackage{subcaption}
\begin{document}

\title{A Bayesian spatial temporal mixtures approach to kinetic parametric images in dynamic positron emission tomography }
% \author{Chuang}
% \author{W. Zhu,    J. Ouyang,
% Y. Rakvongthai, G. El Fakhri and  Y. Fan \footnote{W. Zhu and Y. Fan are with School of Mathematics and Statistics, UNSW Australia, Sydney 2052 Australia\, J. Ouyang and El Fakhri are with Center for Advanced Medical Imaging Sciences, Massachusetts General Hospital, Boston, MA 02114 and Harvard Medical School, Boston, MA 02115, USA\,
% Y. Rakvongthai is with the Department of Radiology, Faculty of Medicine, Chulalongkorn University, Bangkok 10330, Thailand, the work was performed while the author was
% with the Center for Advanced Medical Imaging Sciences, Massachusetts General Hospital, Boston, MA 02114, USA and Harvard Medical School, Boston, MA 02115, USA. }
% }

\author{W. Zhu \footnote{School of Mathematics and Statistics, UNSW Australia, Sydney 2052, Australia}, J. Ouyang\footnote{Center for Advanced Medical Imaging Sciences, Massachusetts General Hospital, Boston, MA 02114 and Department of Radiology, Harvard Medical School, Boston, MA 02115, USA}, Y. Rakvongthai\footnote{Department of Radiology, Faculty of Medicine, Chulalongkorn University, Bangkok 10330, Thailand}\\N. J. Guehl,D. W. Wooten,G. El Fakhri, M. D. Normandin\footnote{Center for Advanced Medical Imaging Sciences, Massachusetts General Hospital, Boston, MA 02114 and Department of Radiology, Harvard Medical School, Boston, MA 02115, USA}\\ Y. Fan\footnote{School of Mathematics and Statistics, UNSW Australia, Sydney 2052, Australia}}

\maketitle

%\begin{abstract}
%We present a fully Bayesian statistical approach to the problem of compartmental modelling in the context of Positron Emission Tomography. We use a mixture modelling approach, incorporating both spatial and temporal information based on reconstructed dynamic PET image. The model is able to borrow information from nearby voxels, allowing for a more robust parameter estimation. Our modelling approach is flexible, and provides uncertainty estimates for the estimated kinetic parameters, which is particularly meaningful for noisy data. Crucially, the proposed method allows us to determine the unknown number of mixture components, which has a great impact on resulting estimated kinetic parameters. We demonstrate our method on simulated dynamic myocardial PET data, and show that our method is superior to both the standard curve-fitting approach and the more recently proposed spatial temporal methods.
%\end{abstract}
%
\begin{abstract}

\textbf{Purpose:} Estimation of parametric maps is challenging for kinetic models in dynamic positron emission tomography. Since voxel kinetics tend to be spatially contiguous, the authors consider groups of homogeneous voxels together. The authors propose {a novel} algorithm to identify the {groups} and estimate kinetic parameters {simultaneously}. Uncertainty estimates for kinetic parameters are also {obtained}. \\
\textbf{Methods:} Mixture {models were} used to fit the time activity curves. In order to borrow information from {spatially nearby voxels, the Potts model was adopted. A spatial temporal model was built incorporating both spatial and temporal information in the data. Markov chain Monte Carlo was used to carry out parameter estimation. Evaluation and comparisons with existing methods were carried out on cardiac studies using both simulated data sets and a pig study data. One-compartment kinetic modelling was used, in which $K_1$ is the parameter of interest,  providing a measure of local perfusion.} \\
\textbf{Results:} {Based on simulation experiments, 
{the median standard deviation across all image voxels, of $K_1$ estimates were 0, 0.13, and 0.16} for {the proposed} spatial mixture models (SMMs), standard curve fitting and spatial $K$-means  methods respectively. The corresponding median mean squared biases for $K_1$ were 0.04, 0.06 and 0.06 for abnormal region of interest(ROI); 0.03, 0.03 and 0.04 for normal ROI; and 0.007, 0.02 and 0.05 for the noise region.}\\
\textbf{Conclusions:} SMM is a fully Bayesian algorithm which determines the optimal number of {homogeneous voxel groups, voxel group membership, parameter estimation and parameter uncertainty estimation simultaneously. The voxel membership can also be used for classification purposes. By borrowing information from spatially nearby voxels, SMM substantially reduces the variability of parameter estimates}.  In some ROIs, SMM also reduces {mean} {squared bias}.
\end{abstract}

{\bf keywords} {PET image, kinetic model, myocardium, spatial mixture model, MCMC.}
\maketitle
\section{Introduction}

Dynamic positron emission tomography (PET) can be used to measure tracer kinetics in vivo, from which physiological parameters, such as tissue perfusion, ligand receptor binding potential, {and } metabolic rate can be determined using compartmental modelling techniques.

{Estimation of the kinetic parametric images can be extremely challenging, since the data are often very noisy.
Most conventional methods either define a region of interest (ROI) and estimate parameters based on the averages\cite{lammertsma1996simplified,slifstein2008comt,nye2008compartmental}, or in a voxelwise fashion. The former requires the identification of ROI which itself is difficult. The latter fails to utilize information from nearby voxels, resulting in more noisy estimates.
Estimations are typically carried out using a minimum least-squares approach \cite{gunn1997parametric}  or a basis function approach\cite{gunn2002positron}. Given low signal-to-noise ratio (SNR), particularly in voxelwise estimations,  some external constraints are often necessary to stabilize parameter estimation. Smoothness regularization can be used\cite{kamasak2005direct,huang1998spatially}, constraining the parameters from nearby spatial locations to be more similar. Similarly, 
Tikhonov regularization \cite{o1999use} can be used to directly constrain parameter values {to be} within a certain range, so that estimates obtained are less sensitive to noise. To account for irregularities in the noise distribution, mixture models can be fitted to each voxel \cite{lin2014sparsity}.  In this case, it is necessary to restrict the total number of mixture components to be small and employ regularization to constrain parameter estimates.}

Assuming a Gaussian error distribution, Bayesian {methods \cite{zhou13} provide an alternative } way of obtaining uncertainty estimates for the kinetic parameters, as well as {model choice for the competing} compartmental models. However, these methods yield {higher voxel to voxel variability} because each voxel was processed independently, and the assumption of Gaussian distribution can also be inappropriate, leading to biased parameter estimates. This has led to the development of several approaches, in which a clustering method was performed first to cluster the PET images into several homogeneous regions, and kinetic parameter estimations were then performed afterward based on the averaged values of each cluster. 
{See, for example, hierarchical clustering of the time activity curves (TACs)}  using a weighted dissimilarity measure\cite{guo03},
 and a comparison of a number of different hierarchical clustering algorithms in this context\cite{velamuru05}.

Recently,  simultaneous clustering and parameter estimation {methods have been proposed using a} spatially regularized $K$-means algorithm. {The algorithm} iteratively estimates the kinetic parameters in a least squared sense between each cluster update\cite{saad2007simultaneous}. {It was } demonstrated that incorporating the physiological model in the clustering {procedure} performed better than their counterparts {in terms of clustering}. However,  the method offers no guidance on the choice of cluster numbers, or how to select the spatial regularization parameter, both can have great influence on the results.  A similar algorithm {was} proposed \cite{mohy2014parametric}, where the clustering and parameter estimation were performed {simultaneously}, although {spatial correlation was ignored}. They demonstrated improvements to parameter estimation in myocardial perfusion PET imaging.

We develop a fully Bayesian approach, based on defining a finite mixture of multivariate Gaussian distributions to model each voxelwise TAC. {We consider that there are a number of distinct homogeneous groups of voxel kinetics which tend to be spatially contiguous.}
The optimal number of mixture components {(or groups)} is estimated via information theoretic {criteria}.
This provides a flexible specification of the error distribution for the TAC. Additionally, we model the spatial dependence between the TACs via the Markov random field {(MRF)}, which allows us to borrow information across nearby voxels. Our model simultaneously handles both spatial and temporal information, making full use of the data available, and this is done with the estimation of the kinetic parameters in a single step.

We apply our approach to simulated one-compartment PET perfusion data and compare the performance of our approach with both the standard voxelwise curve-fitting approach and the spatial temporal approach  \cite{saad2007simultaneous}, using the true kinetic parameters as the gold standard. 
%In this work,
% we assume blood samples are drawn at appropriate sampling frequency and used as the input to the model.
%the input function used was based on a previous 18F-flurpiridaz study, and derived using factor analysis.
 {  We also apply our method to an in vivo pig study data. }

\section{Materials and Methods}
\label{sec:materials}

\subsection{Simulated Dynamic Cardiac Perfusion PET Data}
All the simulation studies were performed using an NCAT torso phantom \cite{segars00} which consists of heart, lungs, liver, and soft-tissue compartments. The  left ventricle (LV) myocardium was segmented into 17 standard segments.\cite{cerwdjkl02}  
{The simulation was based on $^{18}$F-flurpiridaz, which is a new myocardial perfusion tracer that exhibits rapid uptake and longer washout in cardiomyocytes. }
Based on the one-tissue compartmental model, {the TAC of the tissue concentration,} $C_t (t)$, was simulated using
\begin{equation}\label{eqn:TAC}
C_t (t)=K_1 [C_p (t) \otimes \exp (-k_2 t)],
\end{equation}
where  $C_p (t)$ is the blood input function, $K_1$ and $k_2$ are kinetic rate constants for the segment, and $\otimes$ denotes convolution operation. 
{The input function used in the simulation was based on a previously published  $^{18}$F-flurpiridaz study\cite{alpertde12}. During the study, the LV input function was extracted with generalized factor analysis on dynamic series\cite{elfakhri05, elfakhri06}.
This LV input function was treated as the plasma input function. }

\begin{figure}[htp]%Fig1
  \centering
{\includegraphics[width=0.8\linewidth,height=0.25\textheight]{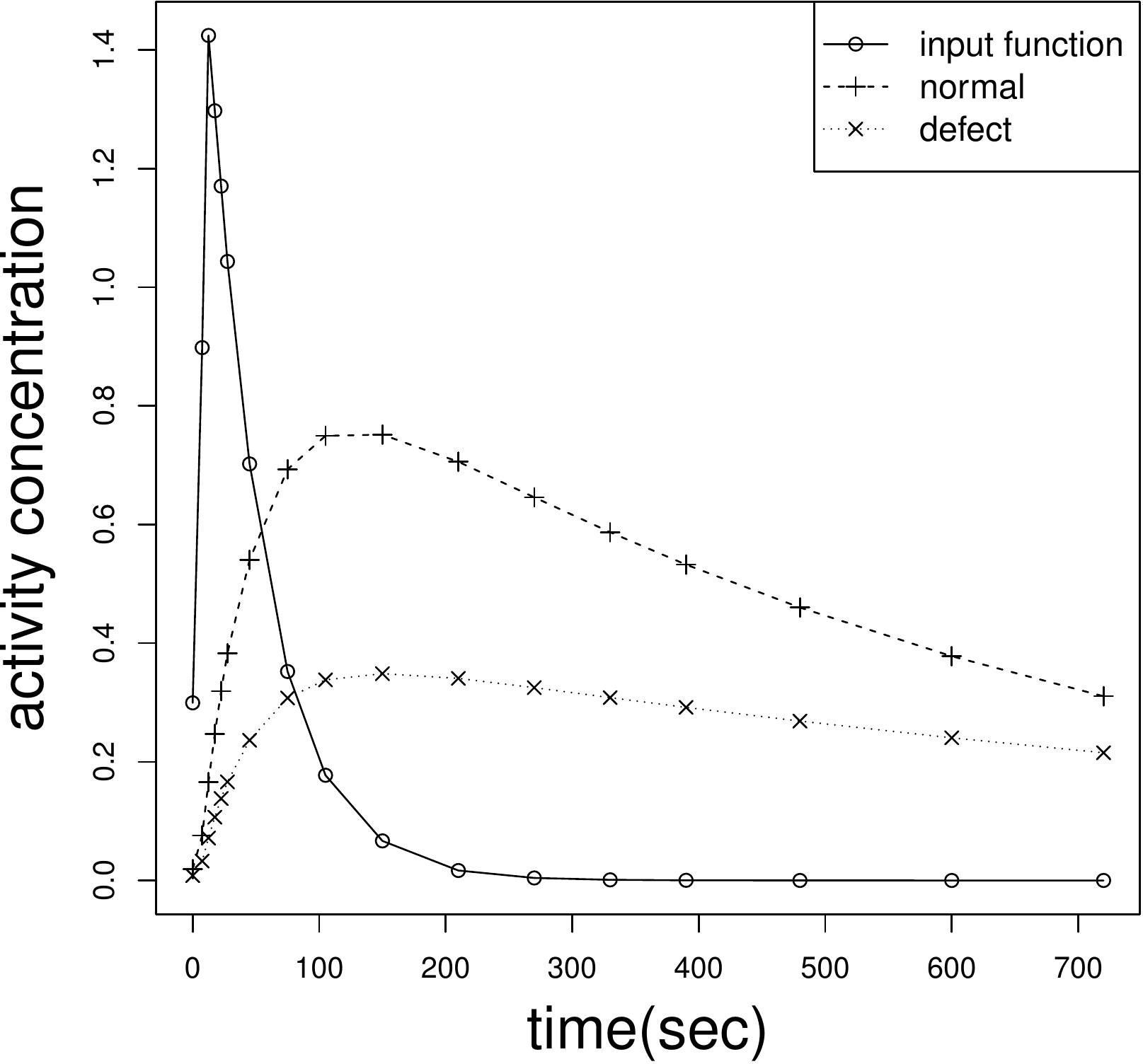}} \\
%\subfigure[]
 \caption{  The input function and two TACs (one normal and one defect segment).}
  \label{inputsample}
 \end{figure}

The kinetic parameters, i.e., ${\boldsymbol k}= (K_1,  k_2)$, assigned to 17 segments were based on the realistic values obtained from PET perfusion studies on normal patients. \cite{alpertde12} In order to mimic a myocardial defect, the segment located in the anterior wall was assigned with values by lowering $K_1$ and $k_2$ by 50\% and 20\%, respectively, of their original values. We added the $18{th}$ segment  to include other voxels not part of the left ventricle myocardium.   Table \ref{segmentname} shows the kinetic parameters assigned to all the 18 segments in the myocardium.  The blood input function $C_p(t)$ and TACs for one normal (basal inferoseptal) and one defect (apex) segments are shown in Figure \ref{inputsample}.

\begin{table} [hbt]
 \centering
  \begin{tabular}{|l|c|c|l|c|c|}
  \hline
  \hline
     segment& $K_1$ &$k_2$ &   segment & $K_1$ & $k_2$ \\
    \hline
    Basal anterior& 0.3665 & 0.0627  & Midinferior  & 0.7162 & 0.0799\\
    \hline
    Basal anteroseptal& 0.6730 &0.0740  & Midinferolateral & 0.8013& 0.0997\\
    \hline
    Basal inferospetal&  0.7656&0.0983  & Midanterolateral &0.7720 & 0.0861\\
    \hline
    Basal inferior& 0.7487 &0.0635 &  Apical anterior & 0.3653& 0.0673\\
    \hline
    Basal inferolateral& 0.9655 & 0.1032  & Apical septal &0.8000 & 0.0861\\
    \hline
    Basal anterolateral& 0.8021 &0.0667  & Apical inferior & 0.7544& 0.0717\\
    \hline
    Midanterior & 0.3438 & 0.0541 &  Apical lateral & 0.6816 & 0.1044\\
    \hline
    Midanteroseptal & 0.7799 &  0.0877 & Apex &0.3290 & 0.0554\\
    \hline
     Midinferoseptal& 0.9016 & 0.0730& Others & 0.7630&0.0820\\
    \hline
    \hline
  \end{tabular}
  \caption{ Segment names and their assigned $K_1$ values {in mL/min/cc, $k_2$ values in 1/min } (i.e.,the ground truth). }
  \label{segmentname}
\end{table}
A system matrix corresponding to  {Philips Gemini PET-CT} camera, which includes position dependent point spread function modelling, a forward-projection operator implemented using Siddon's method, line of response (LOR) normalization factors, and attenuation correction factors, was used to create noise-free sinograms from TACs. \cite{petibonozhrcle13} 
The simulated sinogram data is equivalent to a 13-min dynamic PET scan with the framing scheme of 6 $\times$ 5s, 3 $\times$ 30s, 5 $\times$ 60s, and 3 $\times$ 120s frames.  Twenty five dynamic PET noise realizations were generated. 
%A system matrix corresponding to Siemens whole-body Biograph PET camera, which includes position dependent point spread function modelling, a forward-projection operator implemented using Siddon's method, line of response (LOR) normalization factors, and attenuation correction factors, was used to create noise-free sinograms from TACs. \cite{petibonozhrcle13} 
Both random and scatter events were not included in this study. 
{The total number of events simulated in all the time frames is 50 M. The decay of the tracer was not simulated.
Poisson noise was then added to each pixel in the sinogram based on the mean counts for the pixel.}
%Poisson noise was then added to each sinogram bin.
For each noise realization, the image reconstruction at each time frame was performed using standard ordered subset expectation maximization  \cite{hudson1994accelerated}(OSEM) with 16 subsets and 8 iterations. {No postreconstruction smoothing was applied.} {The physical dimension in the
image reconstruction {was}
57.6cm $\times$ 57.6cm $\times$ 16.2cm, matrix dimension {was} 128 $\times$ 128 $\times$ 36, where the voxel size {was} 0.45cm $\times$ 0.45cm $\times$ 0.45cm.}

{
\subsection{In-Vivo Pig Study Data}
A pig with a body weight of 40 kg was scanned on a Siemens Biograph TruePoint PET/CT with the radiotracer $^{18}$F-flurpiridaz. First, a planar x-ray topogram was performed to allow delineation of the field of view (FOV) and centering on the heart following CT and PET acquisitions. 
The cardiac CT was used for structure localization and later for attenuation correction during reconstruction of PET images. Emission PET data were acquired in 3D list mode and started concomitantly to the injection of $^{18}$F-flurpiridaz, the injected activity was 11 mCI at the time of injection. List mode data were framed into dynamic series of 12 x 5, 8 x 15, 4 x 30, 5 x 60s. PET images were reconstructed using filtered back projection with minimal filtering (voxel size: 2.14x2.14x3 mm3, 55 slices). Attenuation correction was obtained from the CT images. Decay correction was applied and the first 10 min of {the} data are used for kinetic analysis. The input functions for the left and right ventricle were obtained by averaging the TACs from a manually defined region. A one-compartment model with spill-over correction  {was used}. The described experiment was performed under a protocol approved by the Institutional Animal Care and Use Committee at the Massachusetts 
General Hospital.}

\subsection{Kinetic Parameter Estimation Using Curve Fitting}
Kinetic analysis is performed by curve-fitting the TAC in each voxel using a nonlinear least-square fitting,

\begin{equation}\label{eqn:SCF}
\boldsymbol{k}_i= \mbox{argmin}\sum_{t=1}^T w^t(y_i^t-x^t_i({\boldsymbol k}_i))^2,
\end{equation}
where $y^t_i$  is the reconstructed activity concentration for voxel $i$ {at time frame $t$ divided by frame duration $\Delta\tau_t = \tau_{t,e}-\tau_{t,s}$,}   $x^t_i ({\boldsymbol k}_i )=\frac{1}{\Delta {\tau_t}} \int{_{\tau_{t,s}}^{\tau_{t,e} }K_{1,i}[\hat{C}_p(s)\otimes\exp(-k_{2,i}s)]}  ds$, is the average concentration over time frame $t$ {using the current estimates of the kinetic parameters ${\boldsymbol k}_i{=(K_{1,i}, k_{2,i})}$ in voxel $i$}, 
{and measured blood input function $\hat{C}_p(t)$}, $w^t$ is the weighting factor which herein is chosen to be the squared frame duration divided by the total counts in that frame \cite{gunn1997parametric}.  This nonlinear least-square problem can be solved using the Levenberg-Marquardt algorithm. \cite{wangq09} We denote this standard curve-fitting (SCF) approach  in this paper.

\subsection{Spatial $K$-means(SKMS)}
{The spatial SKMS method  performs spatial $K$-means clustering and parameter estimation iteratively \cite{saad2007simultaneous}.}
 The process is as follows. (1) Initialize the cluster means $\mu_ {g},{g}=1,\ldots G$ for a predetermined number of clusters $G$.  (2) {For each $g$, estimate kinetic parameters 
 ${\boldsymbol  k}_g =$ argmin$\sum_{t=1}^T (\mu_g(t)-C_t(t,{\boldsymbol  k}_g))^2$; subject to positivity constraints on ${\boldsymbol  k}_g$. (3) For each voxel indexed by $i=1,\ldots, n$,  reassign cluster membership by minimizing the objective function $\sum_{i=1}^n(\sum_{g=1}^G||y_i - C_t(t,{\boldsymbol  k}_g)||^2 ) + \beta \sum_{r=1}^R I(y_i, y_r)$.  $y_i$ is the TAC at voxel $i$.  $\beta$ determines the influence of the spatial regularizer. $I(\cdot)$ is the indicator function returning a one if $y_i$ and $y_r$ belong to the same cluster and zero otherwise, for all $y_r$ in the neighbouhood of $y_i$. } (4) Based on the new clusters, {calculate $\mu_{g}$ as the mean for each cluster}.
(5) Repeat above steps until {there are no significant changes  in $C_t(t,{\boldsymbol  k}_g)$}. 

There are two main issues for SKMS. First, the authors offered no theoretical guarantee of convergence of their proposed algorithm. Second, {both the number of clusters and spatial regularization parameter $\beta$, } need to be determined but it is not clear {how this can be done. In our implementation of their method, we chose these parameters by looking at a range of $\beta$ and $G$ {values}, and selected the values which minimizes the errors with respect to $K_1$ parameter estimates, setting $\beta$ to 0.2 and $G$ to 17. We note, however, this procedure produces the best possible outcome for SKMS 
but is only possible for simulation data where we know the ground truth. This method was not implemented for the pig study data.}

\subsection{A Bayesian Spatial Mixture Model(SMM)} \label{sec:bsmm}

\subsubsection{Model}
Here, we describe our proposed modelling and estimation approach.
We denote the reconstructed activity concentration data by $ y_i = (y_{i}^1,\ldots, y_{i}^T) \in \mathbb{R}^T$, for voxel $i$. Each data point $y_{i}^t$ corresponds to the reconstructed activity concentration at time $t$. We assume that the data $y_i$ can be grouped into $G$ spatially homogeneous groups, where within each group, all voxels share the same kinetic rate parameters (or TACs) and their variations are only due to noise. The number of groups, $G$, is treated as unknown and { is chosen by the information theoretic model selection criterion, Bayesian information criterion (BIC).}
%We use a mixture of Multivariate Gaussian distribution with $G$ components to model the noisy data. 
%\textcolor{red}{Given the myocardium also contains voxels that have little activity uptake, which are dominated by noise,} we allow one component to cluster this type of voxels. We call this the noise component.
%%Also mixture model allows for a flexible specification of the noise distribution given the fact that the noise distribution in the TAC data is unknown.
%The Potts model \cite{wu1982potts} is used to account for spatial correlation between the  TACs. This is achieved by introducing
%a set of auxiliary random variables  ${\bf z}=(z_1,\ldots, z_n)$, where $z_i$ takes one of the values $1,\ldots, G$, and represents the group/cluster membership for 
%each voxel.  Mathematically, each noisy TAC is given by the mixture of $T$-dimensional Gaussian
%\begin{equation*}
%f(y_i | {\boldsymbol \mu}, {\boldsymbol \Sigma}, \beta)= \sum_{g=1}^G f(z_i=g|\beta)MVN(y_i|z_i=g, {\boldsymbol \mu}_g, {\boldsymbol \Sigma} ),
%\end{equation*}

We use a mixture of multivariate Gaussian distribution with $G$ components to model the noisy data. {Given the ROI used for the analysis may include some voxels outside the  myocardium, as well as some noisy voxels with very little activity uptake inside the myocardium,} we allow one component to cluster these types of voxels. We call this the noise component. The Potts model \cite{wu1982potts} is used to account for spatial correlation between the  TACs. This is achieved by introducing a set of auxiliary random variables  ${\bf z}=(z_1,\ldots, z_n)$, where $z_i$ takes one of the values $1,\ldots, G$, and represents the group/cluster membership for each voxel.  Mathematically, each noisy TAC is given by the mixture of $T$-dimensional Gaussian,\\
\[
f(y_i | {\boldsymbol \mu}, {\boldsymbol \Sigma}, \beta)= \sum_{g=1}^G f(z_i=g|\beta)MVN(y_i|z_i=g, {\boldsymbol \mu}_g, {\boldsymbol \Sigma} ),
\]
where $f(z_i=g|\beta)$ is the marginal density of the Potts model,  $ MVN(y_i|z_i=g, {\boldsymbol \mu}_g, {\boldsymbol \Sigma} )$ is the density of the multivariate 
Gaussian, and $\beta$ is the parameter that reflects the spatial strength between voxels. A value of 0 indicates independence between voxels, while larger values of 
$\beta$ will tend to cluster all voxels into one cluster.
The mean vector of the $g$th multivariate Gaussian component is denoted by ${\boldsymbol \mu}_g$ , and
 ${\boldsymbol \Sigma}  = diag(\sigma^{2,1},\ldots, \sigma^{2,T})$ is the covariance matrix, assumed to be the same for all mixture components. 
 One may relax this assumption to allow more general covariance structure; however, in our simulations studies, the same covariance structure worked well. 
 Here, $\sigma^{2,t}, t=1,\ldots, T$ denotes the variance at time $t$, and the data are assumed to be temporally independent as data were based on the reconstructed 
 image at each time frame. The mixture model representation allows the error distribution to be more flexible. We refer to our model as the SMM.

We set the mean vector for the noise {component} $g^*$ as
\[
{\boldsymbol \mu}_{g^*} = (\mu^1_{g^*},\ldots, \mu_{g^*}^T),
\]
where the $ \mu^t_{g^*}, t=1,\ldots, T$ are unknown parameters.
This component is dominated by noise, taking small values compared with other voxels with larger TAC measurements.
For the remaining components $g=1,\ldots, G-1$, we model the mean vector as a function of the solution to the ordinary differential equation (ODE) describing the one-compartment model \cite{morris04}, although extensions to more
compartments are straightforward.  Hence, for $t=1,\ldots, T$, we set
\begin{equation}\label{eqn:kinetic}
%\mu_g^t=K^g_1 \sum_{\tau=1}^t \hat{C}_p (\tau)\exp(-k^g_2(t-\tau)),
 {\mu_g^t=\frac{1}{\Delta \tau_t} \int_{\tau_{t,s}}^{\tau_{t,e}} K^g_{1}\left[\hat{C}_p(s)\otimes\exp(-k^g_{2}s)\right]  ds},
\end{equation}
where $\hat{C}_p$ is a measured blood input function {and $\Delta \tau_t=\tau_{t,e}-\tau_{t,s}$ is the duration of the $t$th time frame}. The parameters $K^g_1$ and $k^g_2$ are the kinetic rate parameters specific for group $g$.

{For the pig study data analyses, we {modified} Equation \ref{eqn:kinetic} to account for spill-over effects, 
\begin{equation}\label{eqn:kineticSP}
%\mu_g^t=K^g_1 \sum_{\tau=1}^t \hat{C}_p (\tau)\exp(-k^g_2(t-\tau)),
 {\mu_g^t=f_{LV}^g \hat{C}_{LV}(t) + f_{RV}^g \hat{C}_{RV}(t) + (1-f_{LV}^g-f_{RV}^g)\frac{1}{\Delta \tau_t} \int_{\tau_{t,s}}^{\tau_{t,e}} K^g_{1}\left[\hat{C}_p(s)\otimes\exp(-k^g_{2}s)\right]  ds},
\end{equation}
where $f_{LV}^g$ and $f_{RV}^g$ denote the component specific spill-over fractions for the left and right ventricle respectively. 
$\hat{C}_{LV}$ and $\hat{C}_{RV}$ {were} obtained by manually averaging the TACs from the appropriate ROIs.  $\hat{C}_p$ {was} taken as $\hat{C}_{LV}$ multiplied by the plasma fraction, where the plasma concentration ratio was estimated based on blood samples drawn from previous studies. } 

\subsubsection{Prior Specifications}\label{sec:priors}
For Bayesian inference, we need to specify prior distributions for the unknown parameters $K^g_{1}, k^g_{2}$, $\mu_{g^*}^1,\ldots, \mu_{g^*}^T$,  $\sigma^{2,1},\ldots, \sigma^{2,T}, \beta,  g=1, \ldots, G-1$.   We assume independent and uninformative priors for all the parameters, so that the priors are broadly noninformative.

For the kinetic rate parameters, we use the uniform distribution for all $g$,
$ K_{1g} \sim \mathcal{U}(a_{K_1} ,b_{K_1})$ and  $ k_{2g} \sim \mathcal{U}(a_{k_2} , b_{k_2})$,
where $\mathcal{U}$ denotes uniform distribution. We have used $(a_{K_1} , b_{K_1})=(0.3, {\infty})$ and $(a_{k_2} , b_{k_2})={(0, \infty)}$ in our {simulation studies}. {In real applications, one can sometimes get very abnormal rate constants, and a lower value of $a_{K_1}$, such as 0.1 used for our pig study data,  might be appropriate. Setting $a_{K_1}$ much lower than the plausible ranges for $K_1$ will result in additional clusters of the noise voxels being estimated with the kinetic model, and will unnecessarily add to computational cost.} For the mean vector of the noise component $\mu_{g^*}^t \sim \mathcal{U}(0,\infty)$, $t=1,\ldots T$. Setting the prior for $K_1$ sufficiently away from zero allows us to distinguish between the noise component and the non-noise components.
 We set prior $\beta \sim \mathcal{U}(0, b_{\beta})$, where we take $b_{\beta}$ to be 1, so as to include most of the plausible values of $\beta$.
Finally, for the variance parameters $\sigma^{2,t}, t=1,\ldots, T$, we follow the standard approach and use the usual vague conjugate prior with inverse Gamma distribution
$
\sigma^{2,t} \sim IG (a, b)
$,
where $a=0.001, b=0.001$ for an uninformative prior on $\sigma^{2,t}$. {For the pig study data,  we define independent priors for the additional parameters $f_{LV}^g\sim U(0,1)$ and $f_{RV}^g\sim U(0,1), g=1,\ldots,G.$, and set $a_{K_1}=0.1$ and $b_{K_1}=1$.}

\subsubsection{Markov chain Monte Carlo(MCMC)}

Bayesian inference proceeds via  the posterior distribution, obtained by the simple product of the likelihood  and the priors in Section \ref{sec:priors}.
The likelihood function is given by
\begin{equation}\label{eqn:lik2}
 f({\bf y}, {\bf z} |{\boldsymbol \mu}, {\boldsymbol \Sigma}, \beta)= f({\bf z}|\beta) \prod_{i=1}^n f( y_i | z_i, {\boldsymbol \mu}_{z_i}, {\boldsymbol \Sigma}).
\end{equation}

{The posterior distribution is given by the Bayes theorem as the product of the likelihood and the priors 
\begin{equation}\label{eqn:post}
 f( {\bf z}, {\boldsymbol \mu}, {\boldsymbol \Sigma}, \beta |{\bf y})\propto \prod_{i=1}^n f( y_i | z_i, {\boldsymbol \mu}_{z_i}, {\boldsymbol \Sigma})  f({\bf z}|\beta)  f({\boldsymbol \mu}, {\boldsymbol \Sigma}, \beta).
\end{equation}
where the term $f({\boldsymbol \mu}, {\boldsymbol \Sigma}, \beta)$ denotes the prior distribution.}

MCMC algorithms were developed to sample from the posterior distribution, using a combination of random-walk Metropolis-Hastings and Gibbs updates. Details for the implementation of the algorithm {for the model in Equation \ref{eqn:kinetic} are given in the Appendix, the model of Equation \ref{eqn:kineticSP} is a straight forward extension.} Note that occasionally identifiability issues arise in the MCMC estimation of mixtures. Parameters from different components can switch labelling as a result of the invariance of the posterior distribution with respect to labelling. 
This is not an issue when only the MAP estimates are required.  The simplest way to handle this is by imposing certain ordering constraints on parameters \cite{fernandez2002modelling}, or via postprocessing of the MCMC output\cite{zhu2015relabelling}. In this article, the large number of mixture components was adequately handled {using an efficient postprocessing algorithm for MCMC output \cite{cron2011efficient}.}

%\cite{schnatter11} and \cite{cron2011efficient}.
%More relabelling algorithms can be found \cite{zhu2015relabelling}, in which advantages and disadvantages for each relabelling algorithm are presented.
\subsubsection{Determination of the Number of Components G}\label{sec:G}
 One of the {uncertainties} of the above model is the selection of  the value of $G$, which plays a crucial role in the resulting parameter estimation. For model-based inference, in which a likelihood is readily available, a number of  model selection criteria are available, including Bayesian information criterion (BIC), integrated completed likelihood (ICL),  deviance information criterion (DIC) and Akaike information criterion (AIC). The BIC is often considered to be more parsimonious and is the frequently adopted measure of goodness of fit of the model.\cite{steele2009performance} Our approach uses BIC as the criterion to determine the optimal value of $G$.
 % \subsection{Bayesian Information Criterion}

{The} BIC \cite{schwarz1978estimating} is given as
 \begin{equation} \label{bic}
  BIC=-2\log f({\bf y}|G, \hat{{\bf z}}_{MAP},\hat{\theta}_{MAP}) + DF\times (\ln(n)-\ln(2\pi)),
 \end{equation}
where $f({\bf y}|G, \hat{{\bf z}}_{MAP}, \hat{\theta}_{MAP})$ is the likelihood function corresponding to the model with $G$ components, evaluated at the MAP estimator of ${\bf z}$ and  $\theta$, {the vector of all remaining unknown parameters}.  $DF$ is the number of parameters to be estimated, {which includes all the unknown kinetic parameters for each cluster, the variance parameters,  and any hyperpriors which are estimated}. $n$ is the number of observations or voxels. BIC penalizes models with too many parameters against the maximized log-likelihood (or fit to the data). Optimal choice of $G$ corresponds to the model with the smallest BIC value. 

\subsubsection{Implementation}
{We tuned the Gaussian random-walk proposal distributions to obtain an optimal overall acceptance probability of around 20\%-40\%.
For the simulation data sets, we used} $K_{1}^{'g} \sim N(K^g_{1},0.006^{2})$, $k_{2}^{'g} \sim N(k^g_{2},0.001^{2})$, $\mu_{g^*}^{'t} \sim N(\mu_{g^*}^t, 0.00013^2)$  and $\beta^{'} \sim N(\beta,0.002^{2})$. {For the pig data, we used $K_{1}^{'g} \sim N(K^g_{1},0.005^{2})$, $k_{2}^{'g} \sim N(k^g_{2},0.003^{2})$, $\mu_{g^*}^{'t} \sim N(\mu_{g^*}^t, 0.001^2)$,  $f^{'g}_{LV}\sim N(f^g_{LV}, 0.01^2)$, $f^{'g}_{RV}\sim N(f^g_{RV}, 0.01^2)$ and $\beta^{'} \sim N(\beta,0.004^{2}).$}

For a full Bayesian analysis of a single simulated data set,  we ran MCMC for 10000 iterations with the first 4000 iterations discarded as burn-in, and we keep every tenth sample due to high autocorrelation in the MCMC sample.  {Note that for} MAP estimates, {taken as the set of parameter values which gave the highest posterior probability during the MCMC run,} we used {only} 6000 iterations, {as MCMC chains can be expected reach modal regions of the posteriors quite quickly. For the real data set,  8000 iterations of MCMC were obtained with the first 6000 discarded as burn-in}.

We first determine the number of components $G$, by running MCMC for $G=2,\ldots, 26$ and computing the model selection criteria based on BIC (see Section \ref{sec:G}). Then for a fixed $G$, we run posterior inference for a given dataset. For the evaluation of the proposed algorithm, we used 25 replicate simulations. For each replication at the chosen value of $G$, we obtain MAP estimators for comparison with SCF and SKMS. The performance of MAP is known to be worse than the posterior mean estimate, but it is sufficient to provide a good guide on the quality of the inference. The results are presented in Section \ref{sec:results}.

\subsection{Performance Evaluation}
For a given noise realization, $n$, we compute kinetic parameter bias of voxel $i$ using
\begin{equation}\label{eqn:bias}
b^n_i=({\boldsymbol k}^n_i-{\boldsymbol k}^{Tr}_i)/({\boldsymbol k}^{Tr}_i ),
\end{equation}
where  $b^n_i$ is the bias of estimated kinetic parameter using the true kinetic parameter ${\boldsymbol k}^{Tr}_i$ as gold standard.  Based on 25 noise 
realizations, we compute the mean bias $\bar{b}_i$,  {the mean squared bias $\bar{b}^2_i$,}  and standard deviation $s_i$ bias for voxel $i$ using
\begin{equation}\label{eqn:tbs}
\bar{b}_i=\frac{\sum_{n=1}^N b^n_i}{N}, \quad {\bar{b}^2_i=\frac{\sum_{n=1}^N (b^n_i)^2}{N}},\quad s_i=\sqrt{\frac{\sum_{n=1}^N(b^n_i-\bar{b}_i)^2}{N-1}},
%\textcolor{red}{\bar{b}^2_i=\frac{\sum_{n=1}^N (b^n_i)^2}{N}, \quad s_i=\sqrt{\frac{\sum_{n=1}^N(b^n_i- \sum_{i=1}^N b_i^n/N)^2}{N-1}},}
\end{equation}
where $N$ is the total number of noise realizations.  We perform the calculations described above for SCF, SMM and SKMS methods and make a comparison between  methods.

\section{Results} \label{sec:results}

\subsection{Model Selection}
\begin{figure}[ht]%Fig2
  \centering
   \includegraphics[width=0.8\linewidth,height=0.3 \textheight]{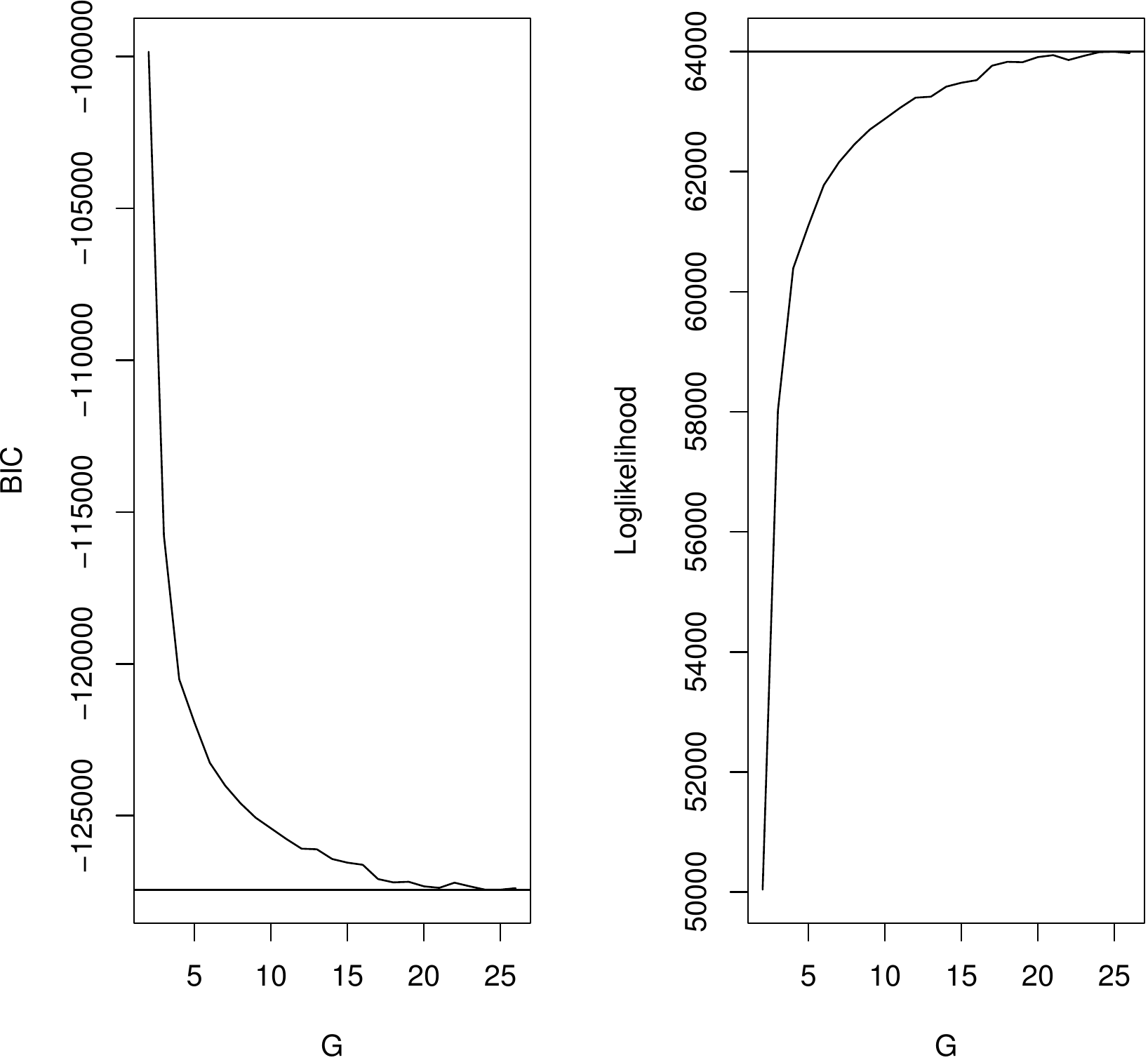}
 \caption{BIC values (left panel) and log likelihood (right panel) for $G=2,3,\ldots,26$ in the spatial mixture model. The horizontal lines indicate the minimum and the maximum of BIC and log likelihood respectively.}
 \label{BICplot}
\end{figure}

Figure \ref{BICplot} shows the BIC values (left panel) and the corresponding log likelihood (right panel) for competing models {for a single noise realization}.  Horizontal lines in both subfigures denote the minimum and maximum values for BIC and  the log likelihood respectively.  The log likelihood values are expected to keep increasing {with $G$, while the BIC penalizes the use of additional parameters in models with larger $G$. }
Both the BIC and log likelihood changed dramatically from $G=2$ to about $G=10$, preferring models with larger $G$ values, and this stabilized after around $G=17$. Part of the changes seen here can be attributed to Monte Carlo errors. Thus, a parsimonious choice for $G$ would be $G=17$, representing the model with 16 TAC components and one noise component. Subsequent results {for simulated data} in this paper were generated by the model with $G=17$. {The same value of $G$ was also found for the pig study data}.

%\subsubsection{Inference of kinetic parameter}
\subsection{Parameter Estimation and Comparison to Existing Methods}
\begin{figure}[ht]%Fig3
  \centering
  \includegraphics[width=14cm, height=18cm, keepaspectratio]{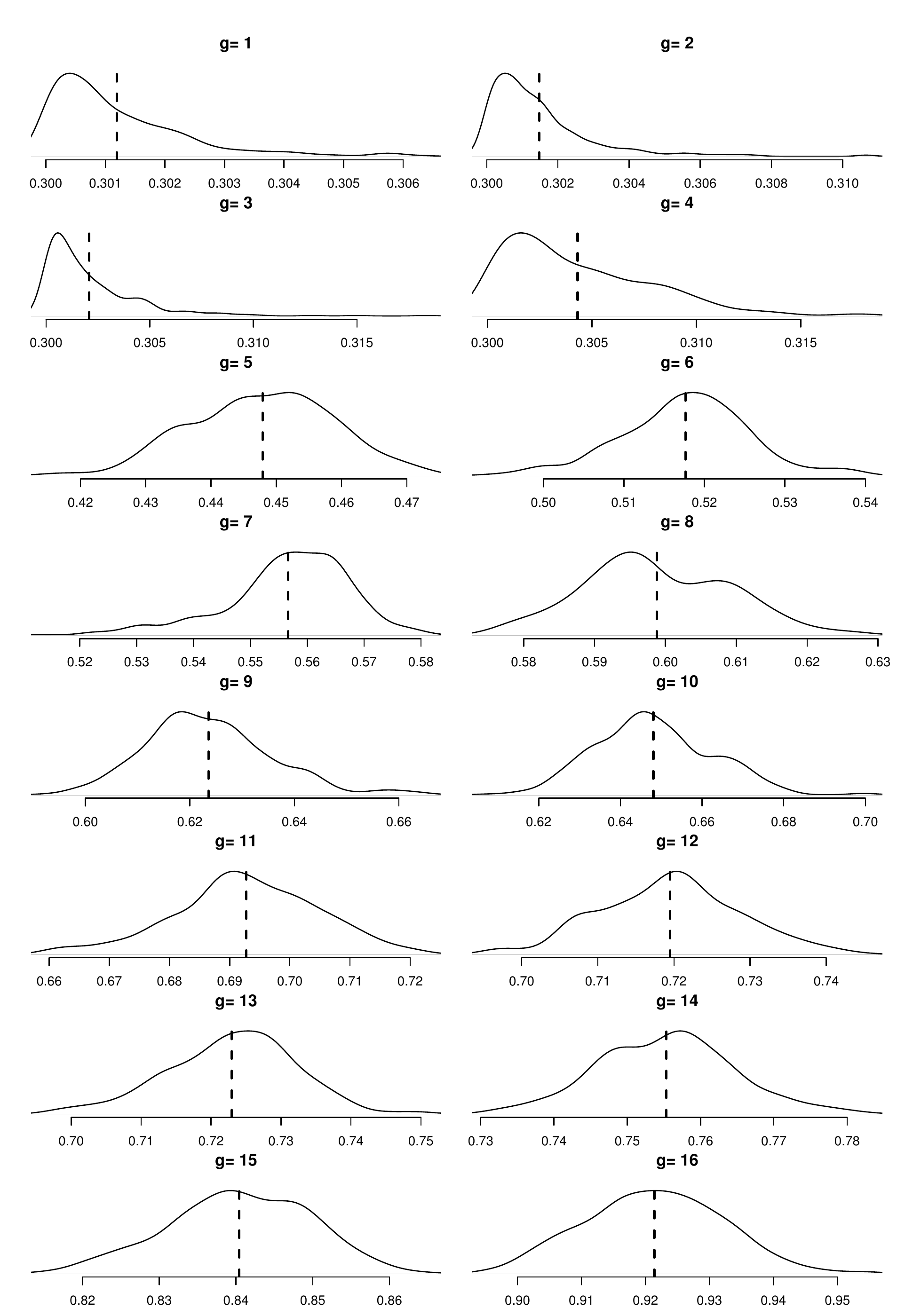}
 \caption{{Marginal} posterior density of $K^g_1$ for $g=1,\ldots, 16$ clusters. Vertical dashed line denotes corresponding posterior means. {Based on a single noise realization of simulation data. }}
  \label{k1plot_g17}
 \end{figure}

\begin{figure}[ht]%Fig4
  \centering
  \includegraphics[width=14cm,height=18cm, keepaspectratio]{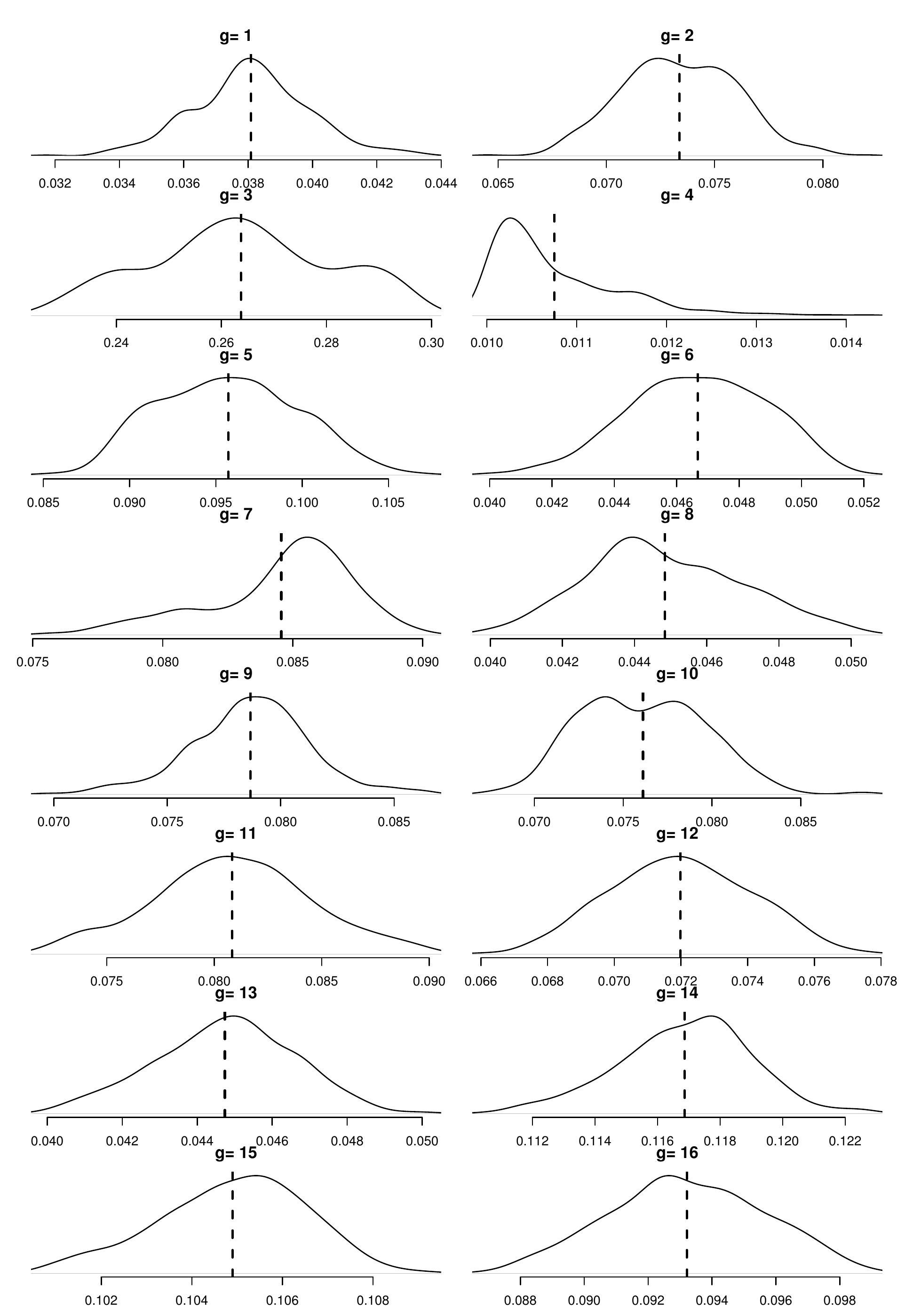}
 \caption{Posterior density of $k^g_2$ for $g=1,\ldots, 16$ clusters. Vertical dashed line denotes corresponding posterior means. 
 {Based on a single noise realization of simulation data.}}
  \label{k2plot_g17}
 \end{figure}

%The mixture model with $G=17$ components represents the model with 16 components with kinetic activity and one noise component.
% Table \ref{parameter} presents the posterior mean and 90\% posterior credibility intervals of the kinetic parameters, i.e., $K_1$ and $k_2$,  for the 16 of the 17 components. 
%Figure \ref{fig:tac} presents the mean TAC for all 16 components other than the noise component, using the posterior mean (Table \ref{parameter}). Superimposed is the averaged observed TAC from the noisy reconstructed image data. All the estimated curves are very close to the  observed curves. This indicates that our method has fitted the data well. 
For the simulation data, Figures \ref{k1plot_g17} and \ref{k2plot_g17}  show the corresponding  {marginal} posterior distributions of $K_1$ and $k_2$ respectively, {with vertical lines indicating the posterior mean, and the uncertainty of the estimates indicated by the spread of the distributions}.

\begin{figure}[htp]%Fig5
  \centering
% \begin{tabular}{cc}

%{\includegraphics[width=0.45\linewidth,height=0.18\textheight]{K1bias_abnorm.eps}}
%{\includegraphics[width=0.45\linewidth,height=0.18\textheight]{k2bias_abnorm.eps}} \\

{\includegraphics[width=0.45\linewidth,height=0.18\textheight]{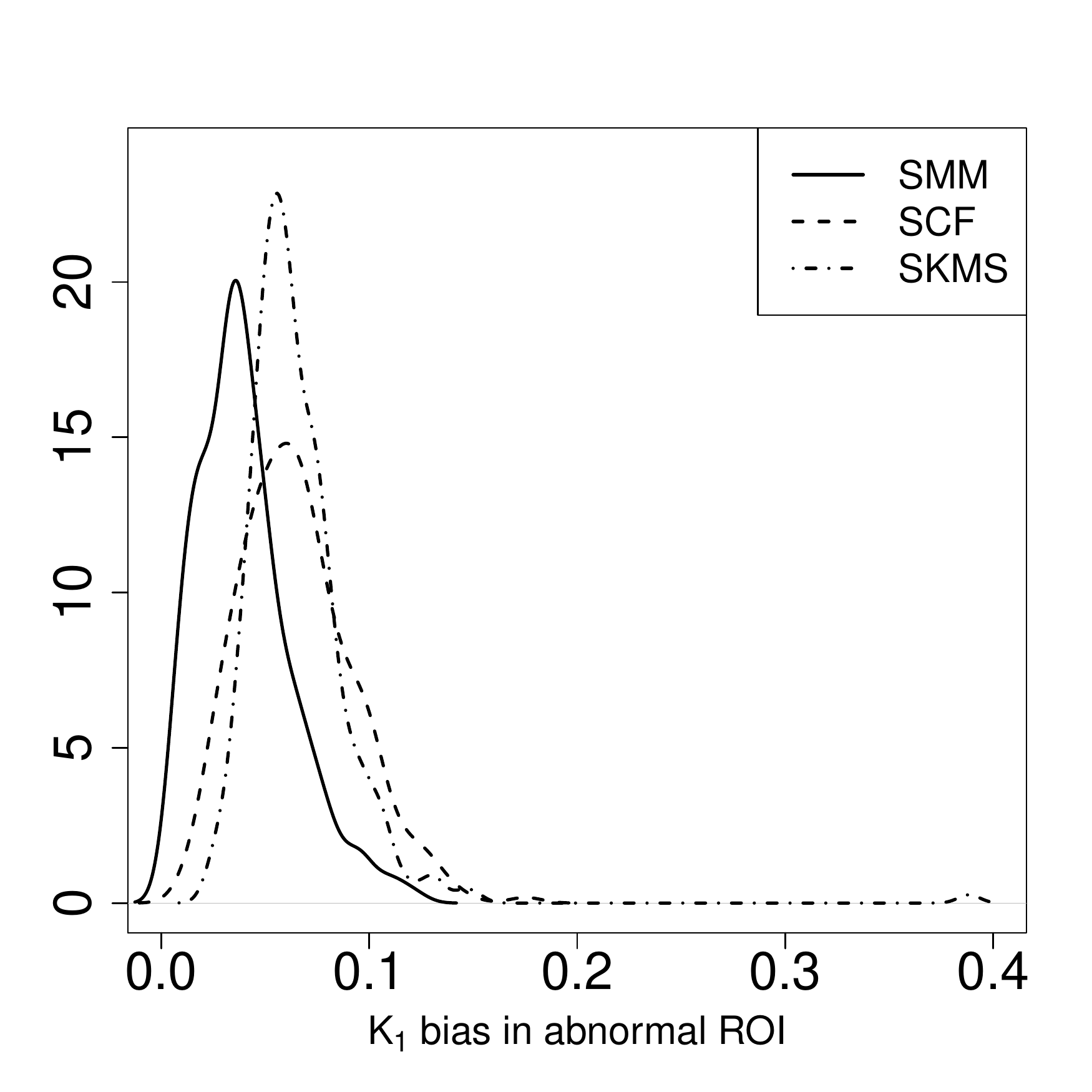}}
{\includegraphics[width=0.45\linewidth,height=0.18\textheight]{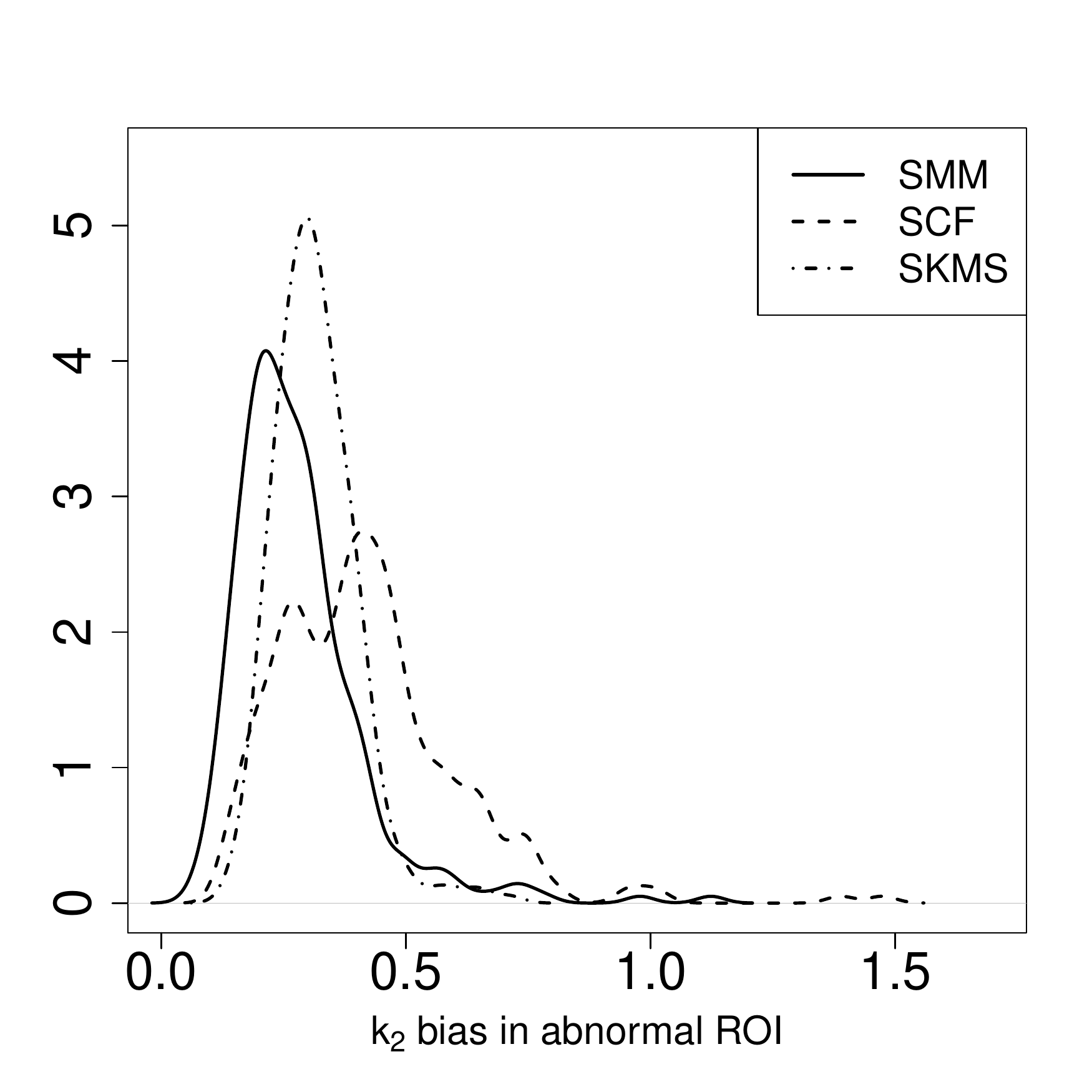}} \\

 %{\includegraphics[width=0.45\linewidth,height=0.18\textheight]{K1bias_norm.eps}} 
 %{\includegraphics[width=0.45\linewidth,height=0.18\textheight]{k2bias_norm.eps}} \\
 
 {\includegraphics[width=0.45\linewidth,height=0.18\textheight]{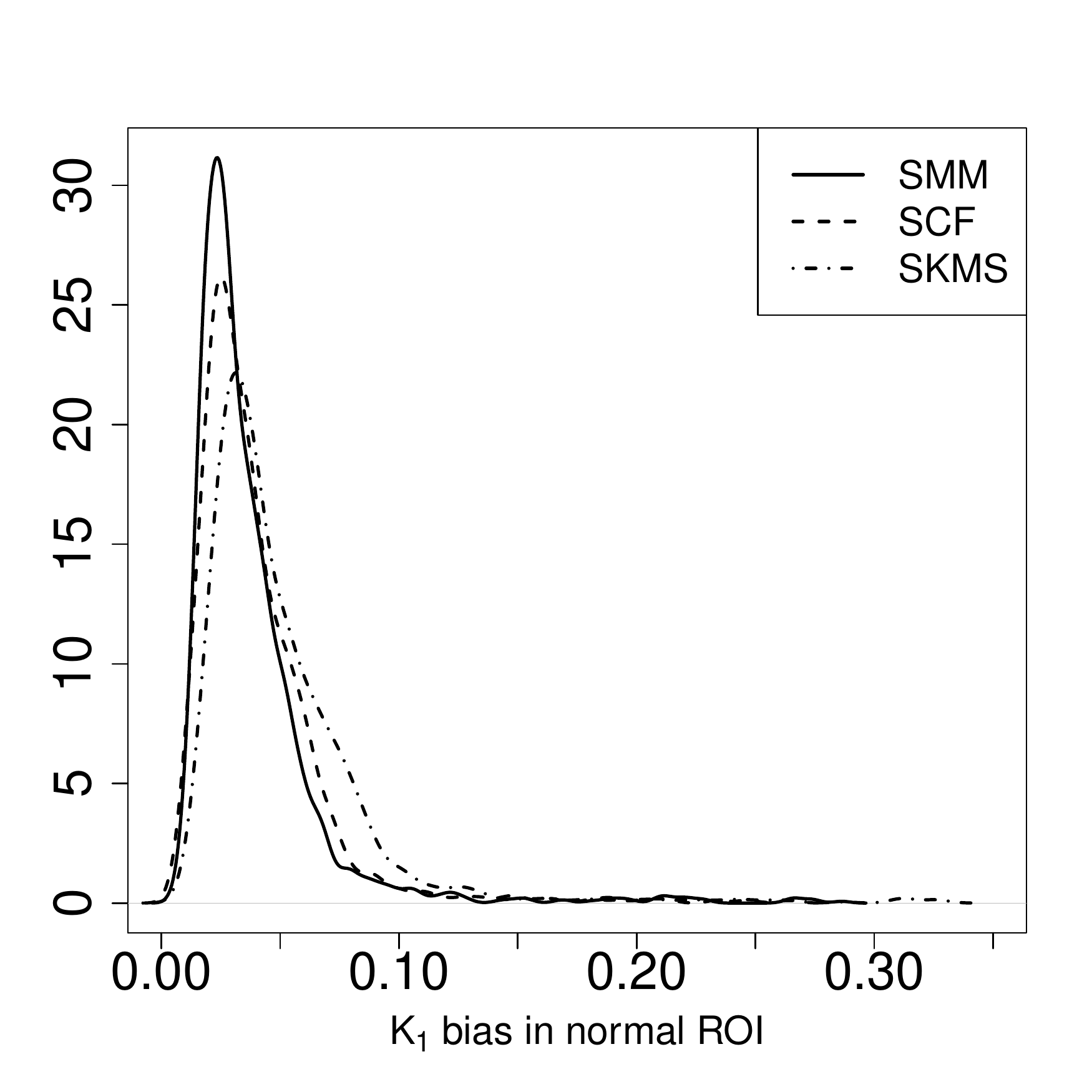}} 
 {\includegraphics[width=0.45\linewidth,height=0.18\textheight]{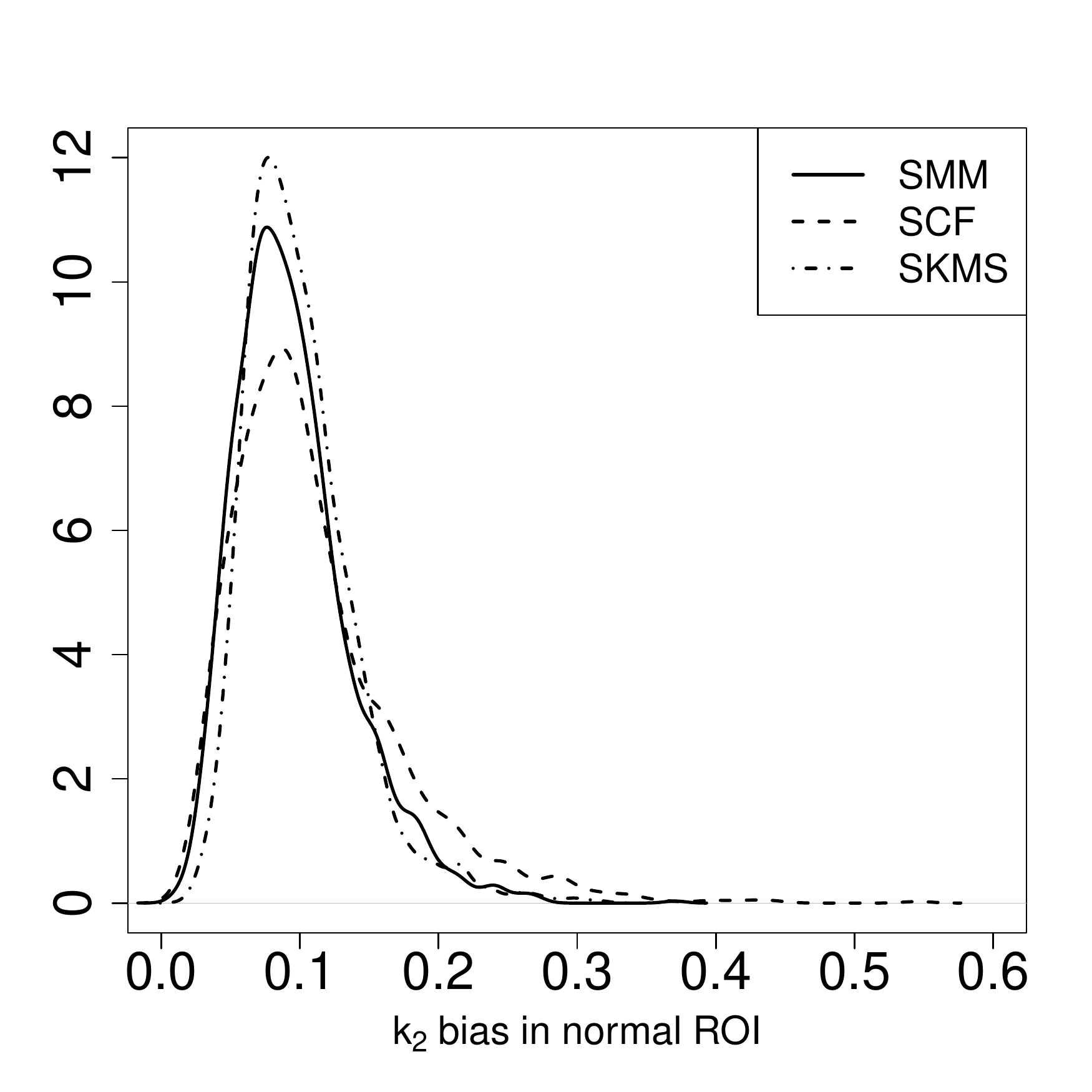}} \\
 
 {\includegraphics[width=0.45\linewidth,height=0.18\textheight]{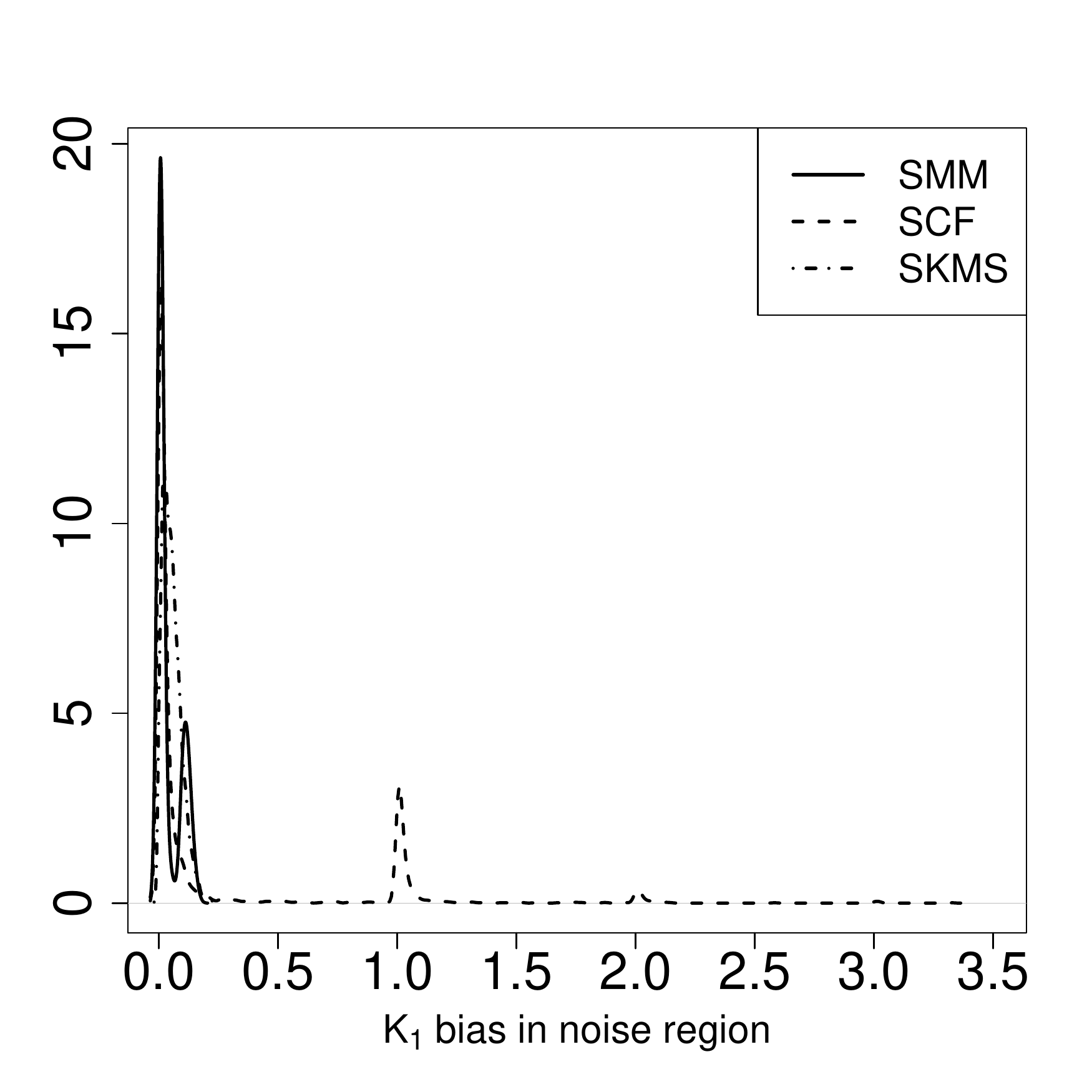}} 
 {\includegraphics[width=0.45\linewidth,height=0.18\textheight]{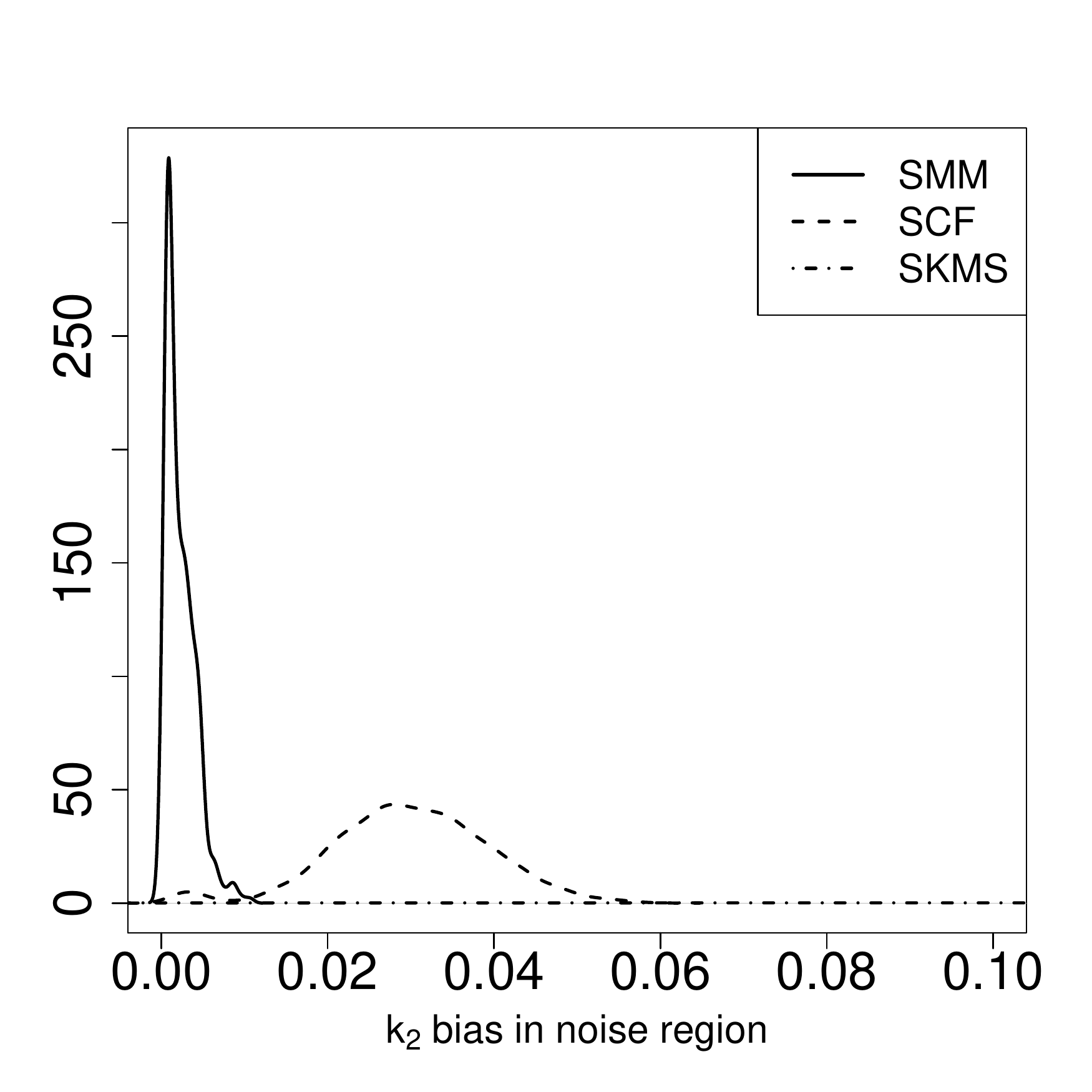}} \\
 
 {\includegraphics[width=0.45\linewidth,height=0.18\textheight]{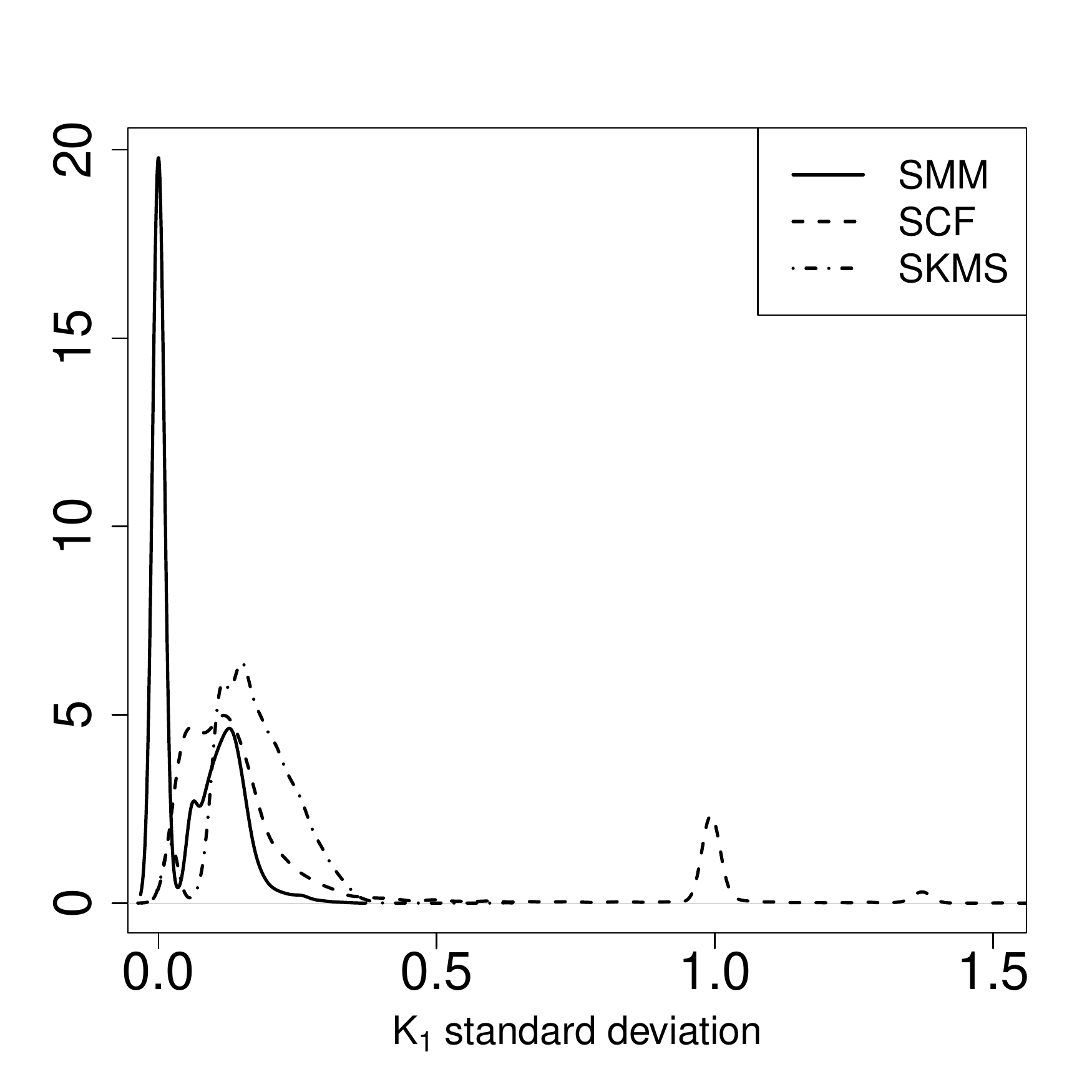}} 
{\includegraphics[width=0.45\linewidth,height=0.18\textheight]{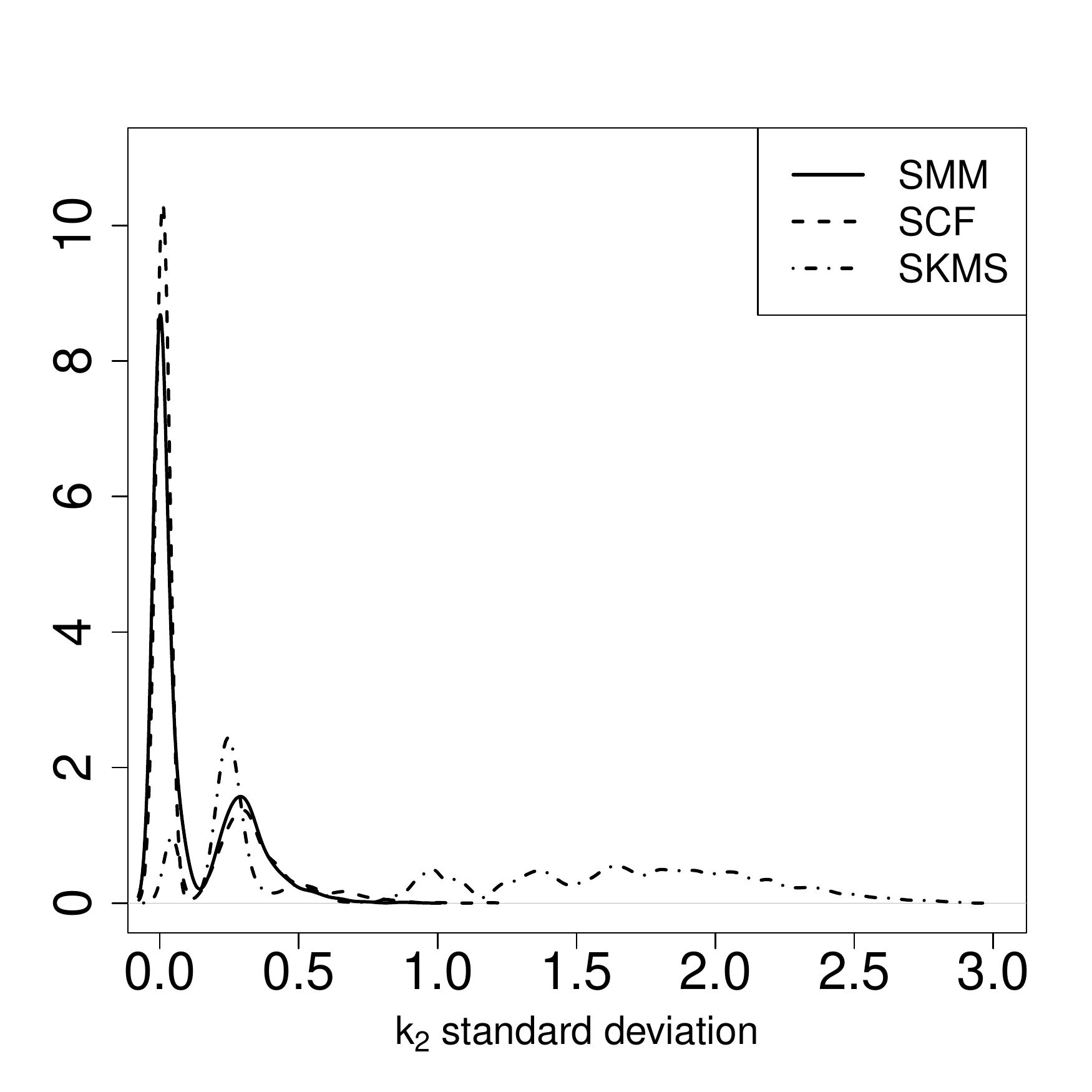}}
% \end{tabular}
 \caption{{Distributions of the mean squared biases and standard deviation of biases for SMM (solid line); SCF (dashed line) and SKMS (dotted and dashed line). The first three rows show the mean squared biases for $K_1$ and $k_2$ in the abnormal, normal, and the noise ROIs respectively.
 The last row shows the standard deviation of the biases. Mean squared biases and the standard deviation of biases are calculated according to Equations \ref{eqn:bias} and \ref{eqn:tbs}, over 25 replicate simulation data sets.}}
  \label{kbias}
 \end{figure}

 To assess the robustness of our estimation procedure and its performance against existing methods, we repeated our estimation procedure for 25 replicate data sets, obtained from the same simulation setup. {We implemented the three competing methods SMM, SCF and SKMS. There was a single extremely large value of $k_2$ estimate from SKMS, which we omit from the results shown.
 Figure \ref{kbias} shows the distribution of the mean squared biases of  $K_1$ and $k_2$ in the abnormal, normal and noise regions, and the overall standard deviation of the biases. The computations were  calculated according to Equation  \ref{eqn:tbs}, with the exception that in the noise region, the bias was computed by setting {the denominator of Equation \ref{eqn:bias}}, $\boldsymbol k_i^{Tr}$, to 1, since we cannot divide by zero. }

{Results for the abnormal ROI are shown in the first row of Figure \ref{kbias}. For $K_1$, the mean squared biases for SMM ranged from 0.006 to 0.12, with a median of 0.04. For SCF, the range was from 0.013 to 0.175, with a median of 0.06. For SKMS, the range was between 0.02 and 0.39, and the median was 0.06. For the $k_2$ estimation, the biases ranged from 0.06 to 1.12 for SMM,  the median was 0.25. For SCF, the range was between 0.13 and 1.47, and the median was 0.4. For SKMS, the biases ranged from 0.14 to 0.7, and the median was 0.31.}

{The normal ROI is shown in the second row of Figure \ref{kbias}. For $K_1$, the mean squared biases for SMM ranged from 0.006 to 0.29, with a median of 0.03. For SCF, the range was 0.005 to 0.27, with a median of 0.03. For SKMS, the range was between 0.007 and 0.33, and the median was 0.04. For the $k_2$ estimation, the biases ranged from 0.01 to 0.37 for SMM, with a median of 0.09. For SCF, the range was between 0.01 to 055, and a median of 0.1. For SKMS, the biases ranged from 0.02 to 0.32, and the median was 0.09.}

{The noise region is shown in the third row of Figure \ref{kbias}. For $K_1$, the mean squared biases for SMM ranged from 0.004 to 0.165, with a median of 0.007. For SCF, the range was 0.0001 to 3.32, with a median of 0.02. For SKMS, the range was between 0.0005 and 0.22, and a median of 0.05. For the $k_2$ estimation, the biases were approximately 0 for SMM. For SCF, the biases ranged between 0 to 0.06, and the median was 0.03. For SKMS, the biases ranged from 0 to 13.05, and the median was 2.96.}

{The last row of Figure \ref{kbias} shows the standard deviations of the biases for $K_1$ and $k_2$. For $K_1$, the standard deviations of biases ranged from 0 to 0.34 for SMM, with a median of 0. For SCF, the range was between 0.008 to 1.70, with a median of 0.13. For SKMS, the range was between 0.073 to 0.63, and the median was 0.16. For the $k_2$ standard deviations of bias, the range was between 0 and 0.93 for SMM, with a median of 0. For SCF, the range was between 0.002 to 1.19, and a median of 0.01. For SKMS, the range was between 0.02 and 2.89, with a median of 1.20.}

\begin{figure}[ht]%Fig6
  \centering
  \begin{subfigure}{0.45\textwidth}
                \includegraphics[width=\textwidth,height=0.24 \textheight]{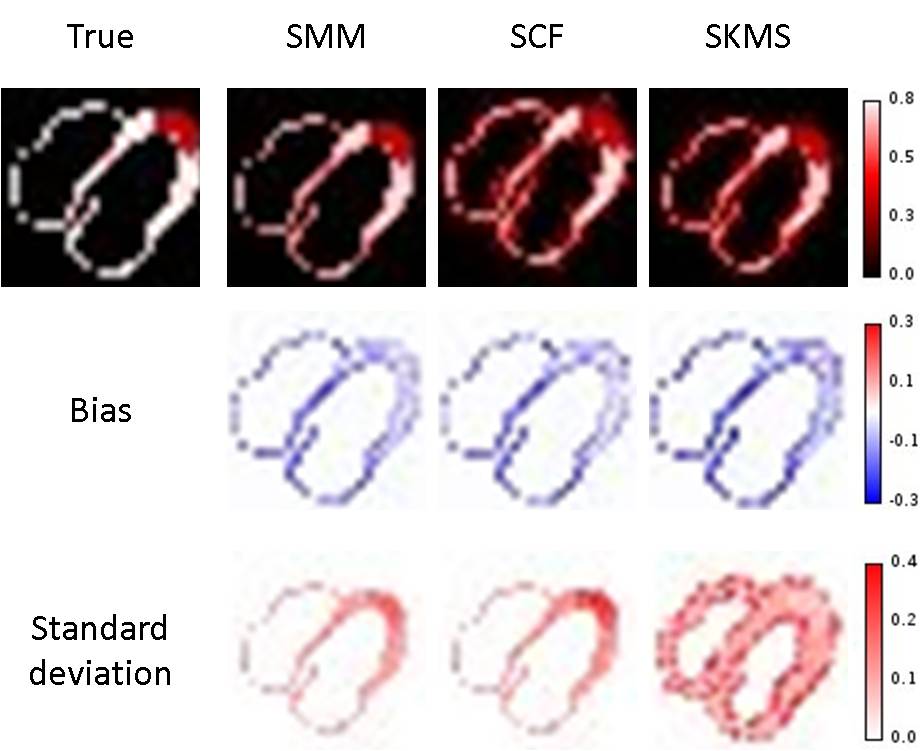}
                \caption{$K_1$ } %estimation, bias and standard deviation.
                \label{k1graph}
 \end{subfigure}
 \begin{subfigure}{0.45\textwidth}
                \includegraphics[width=\textwidth,height=0.24 \textheight]{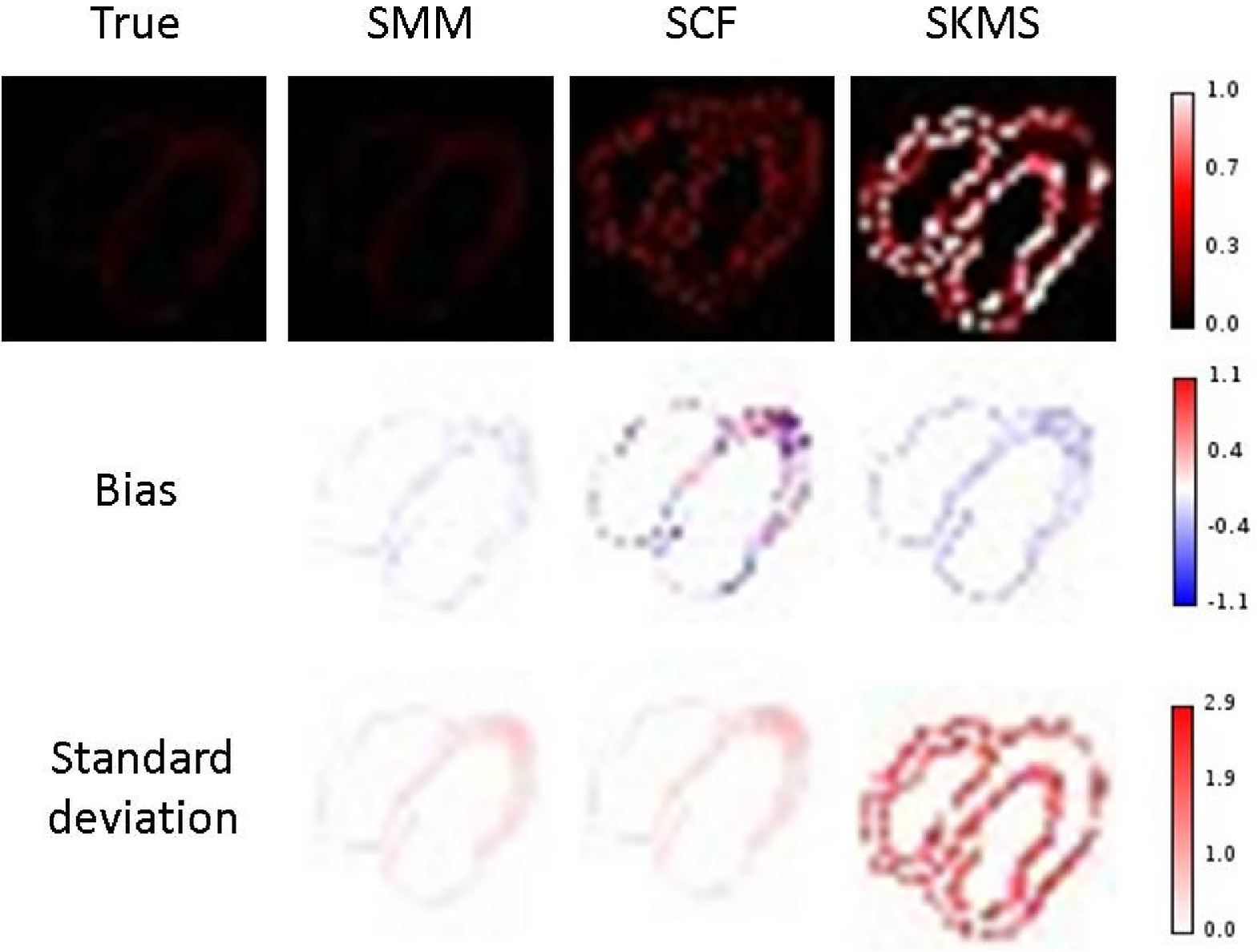}
                \caption{$k_2$ } %estimation, bias and standard deviation.
                \label{k2graph}
 \end{subfigure}
 \caption{Parameter estimates, bias and standard deviation of bias for a single slice of the image. Comparisons for $K_1$ (a) and $k_2$ (b) {for 25 replications of simulation data.}}
  \label{petereu}
\end{figure}

Figure \ref{petereu} compares the bias and standard deviation of bias between SMM, SCF and SKMS for a single slice. The bias is calculated according to the first term in Equation \ref{eqn:tbs}, this is the average of the biases over 25 replications.  Figure \ref{k1graph} shows the $K_1$ estimates.  {The biases ranged from -0.27 to 0.10, -0.28 to 0.09 and -0.33 to 0.08 respectively, 
for SMM, SCF and SKMS. Similarly, the standard deviations ranged from 0 to 0.26, 0 to 0.37 and 0 to 0.35 respectively. For the $k_2$ estimates in Figure \ref{k2graph}, the biases ranged from -0.22 to 0.2, -1 to 1.06 and -0.46 to 0.28 respectively. The standard deviations ranged from 0 to 0.59, 0 to 0.86 and 0 to 2.89 respectively.}

\begin{figure}[htp]%Fig7
  \centering
\includegraphics[width=0.7\textwidth, height=7cm]{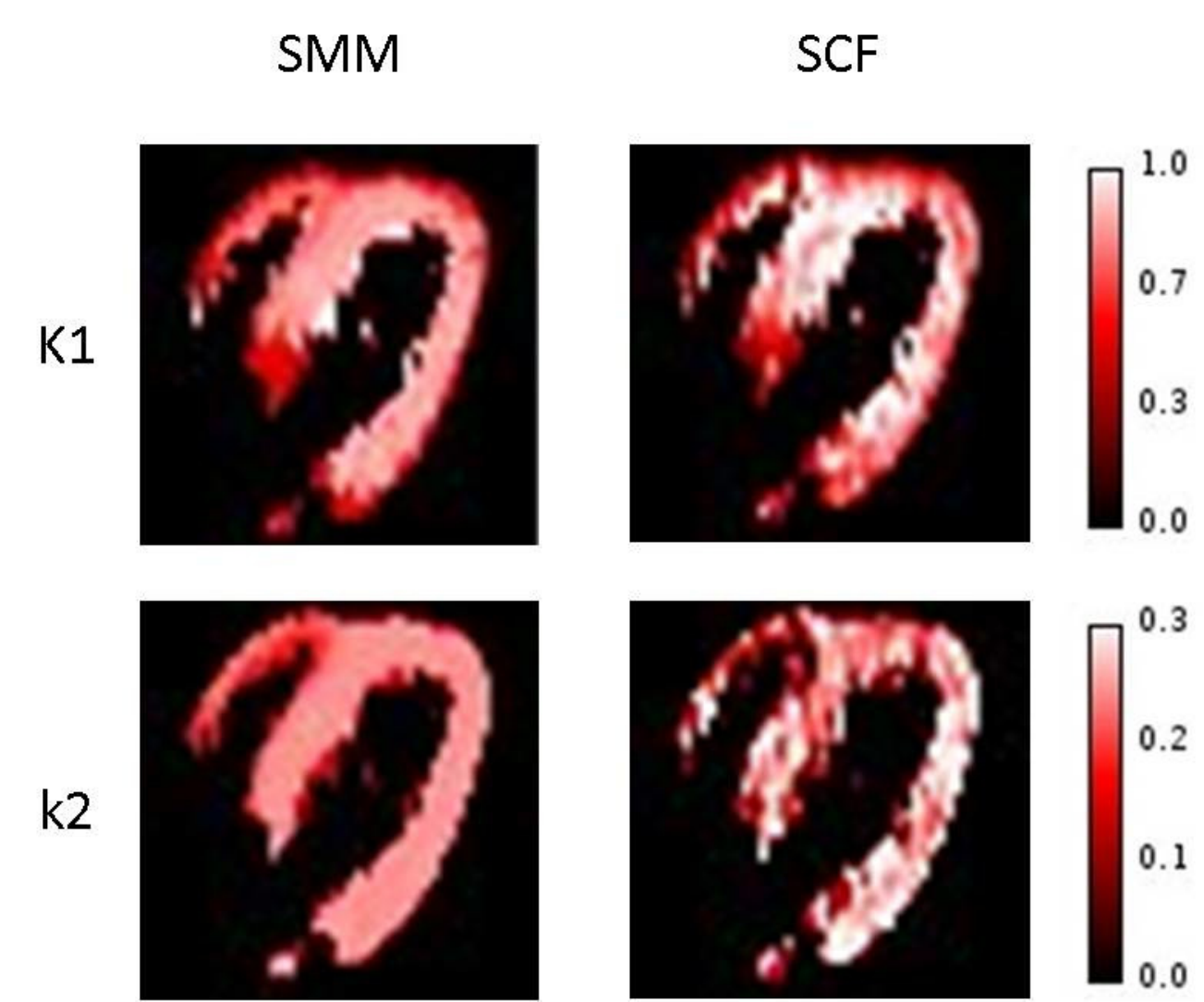}
 \caption{Parameter estimates for a single slice of the pig study data.}
  \label{petereuPig}
\end{figure}

{Figure \ref{petereuPig} compares a single slice of the kinetic parametric images between SMM and SCF for the pig study. $K_1$ parameters were constrained to be between $0$ and $1$ in both SMM and SCF estimation, as unconstrained estimation lead to many physiologically implausible large values of $K_1$}.

\section{Discussions}

This paper proposes a novel method, SMM, that clusters voxelwise TACs and estimates kinetic parameters simultaneously.
Our modelling approach shares similarities to the recently proposed work \cite{lin2014sparsity}, where the mixture model {was fitted} to each voxel (while still borrowing information across nearby voxels) to overcome the issue of non-Gaussian error distributions.  We allow several similar voxels to share the same parameter values, since separate mixture models fitted to each voxel introduces too many parameters, {and will lead to more estimation uncertainty}. Our approach naturally allows us to constrain parameter estimates without the need to specify regularization parameters as in the usual Bayesian maximum a posterior (MAP) approaches. Finally, we allow the data to determine the most appropriate number of mixtures to fit to the data. 

Our model-based approach offers several advantages,  compared to other existing statistical approaches described above.  We require minimal user input in the algorithm, preferring to allow the data to dictate the optimal choices. One benefit of our modelling approach is in the determination of the optimal number of mixture components. We also automatically compute the value of the smoothing parameter used in the MRF model. This unknown parameter is difficult to estimate, and in many applications of spatial modelling, the estimation of this parameter {has} not been carefully considered. The choice of both these parameters can have a big impact on results, {since suboptimal choices will either result in higher bias or higher variance for the resulting parameter estimates}. Finally, the Bayesian statistical framework allows us to quantify uncertainty probabilistically, since uncertainty in the model and parameters is a natural consequence of our modelling framework. An efficient 
MCMC algorithm allows us to provide parameter estimates, as well as uncertainty quantification simultaneously.

{In Figure \ref{BICplot}, the BIC  values start to reach a minimum at around  $G=$16 or 17. We chose to work with 17, but higher values of 18 or 19 will work equally well. These numbers are similar to the number of true segments simulated; however, we expect that this number can be different depending on the nature of the noise. Figures \ref{k1plot_g17} and \ref{k2plot_g17} indicate the component mean estimates for $K_1$ and $k_2$. It is difficult to make direct correspondences between the clusters we {obtained} with the true segments. The first four components correspond mostly to noise,  the next four components correspond mostly to abnormal voxels and the rest belong to normal voxels. The discrepancy between the estimated values of $K_1$ and the truth is most obvious in the abnormal region, this is possibly a combination of partial volume effect, as well as misclassification of the normal voxels. Given that SMM outperforms the other two methods in the abnormal region, we believe similar 
issues 
with the data are affecting the other two methods also. Despite the fact that it is difficult to make sense of individual clusters, aggregating the clusters can provide us with information about the larger ROIs. For instance, if we are interested in identifying the three regions of noise, abnormal and normal, we can aggregate the clusters  according to $K_1<0.3$, $0.3 \leq K_1<0.6$ and $K_1\geq 0.6$ respectively.  A similar procedure can be used to classify the regions using the results from SCF and SKMS. In terms of misclassification rates,  based on a single simulation data set, SMM classified 96.34\% of noise voxels correctly, compared to 94.43\% and 87.60\% for SCF and SKMS. The misclassified voxels for SMM were all
assigned to the abnormal voxels, this corresponds to the first four clusters in Figure \ref{k1plot_g17}. For the other two methods, they were spread between abnormal and normal voxels. For the abnormal region, SMM had a 100\% correct
classification, while this was only 62.68\% for SCF and 52.82\% for SKMS. All of the misclassifications in SCF and SKMS were allocated to noise. Finally, for the normal region, SMM, SCF and SKMS had  69.57\%, 73.49\% and 58.99\% respectively for correct classifications, most of the misclassifications were found to be allocated to the abnormal region.}

%Figure \ref{fig:tac}  allows us to assess visually the goodness of fit of our estimation. Here the observational mean of each cluster is calculated to compare with the estimated TAC using SMM. {The figures show a good agreement for all clusters }.  The adherence to mean of the reconstructed data shows that our model gives very good fit.
% This suggests that there might be some mis-specification of the model we fitted.

{In terms of $k_2$ estimation, SMM is clearly better than the other two methods. This can be seen clearly in Figures \ref{kbias} and \ref{petereu}. SKMS performed the worst, particularly in the abnormal and noise regions, their parameter estimation can be prone to very large biases. In the noise region, in particular, the median mean squared bias was around 2.96, while the other two methods were close to 0. Putting constraints on these parameters may prove useful. }

{For $K_1$ estimates, SKMS was marginally worse than the other two methods in terms of mean squared bias. In the abnormal region, SMM shows noticeably superior performance, where it can be seen in Figure \ref{kbias}, top row, the entire distribution of SMM is closer toward 0 than the other two methods. 
The difference in the normal region is less obvious. The third row in Figure \ref{kbias} shows the biases in the noise region; here, since SMM set $K_1$ in this region to 0, the graph can be interpreted by looking separately at the values of mean squared bias below $(0.3)^2=0.09$ and above. On average, voxels with bias greater than this value are essentially misclassified, i.e., they should be singled out as noise, but instead have significant values for the kinetic parameters. For SMM, there {was} an average misclassification rate of 3.66\%, for SCF it {is} 5.71\% and 
11.65\% for SKMS. SCF has the largest mean squared biases here, going up to 3.32, while the other two methods remain around 0.2.}

{In terms of the standard deviations of the bias, SMM performed the best, while SCF was the worst. The plot in the last row of Figure \ref{kbias} shows that for $K_1$,
the range for SCF goes up to 1.7, while for SMM and SKMS, this was only 0.34 and 0.63 respectively. In fact, 15\% of the voxels estimated by SCF was greater than 0.34 (the largest value obtained by SMM), and 0.8\% from SKMS.}

The proposed method is clearly superior in terms of robustness, indicated by the substantially smaller standard deviation estimates, as can be seen in both Figure  \ref{kbias}  and Figure \ref{petereu}. It also performed at least as good as, and sometimes better than the other two methods in terms of mean squared bias. {In the single
slice plot in Figure \ref{petereu}, where the mean of the raw biases was plotted, it is difficult to distinguish between the three methods. This is due to the fact that when raw biases are averaged, they will go toward zero as the effects of the large positive and negative biases cancel out. This will be true for all unbiased estimators regardless of how sensitive the estimations are to noise.} In this sense, it is more useful to look at the mean squared or absolute biases.

{In terms of the estimation of the pig study data, it was not possible to compare the bias and standard deviations of the biases because the ground truth {was} not known. However, the parametric images
shown in Figure \ref{petereuPig} suggest that much smoother $K_1$ and $k_2$ images were produced by SMM compared to SCF. The white color in the $K_1$ images indicates a value close to 1, which is the upper bound of the artificial constraint we used. It is clear from the figure that many values produced by the SCF 
method were simply truncated at this value. SMM estimation produced significantly less values close to the upper bound.} 
{The upper bound of 1 for $K_1$ is essentially arbitrary. For SMM estimation, if we remove this bound, we obtain two groups of voxels with physiologically implausibly high $K_1$ values whilst the rest of the voxel estimates remains unchanged, well below 1. However, in terms of the SCF estimation, raising the bounds to higher than 1 produced many more voxels between 1 and 2. However, since this was a resting pig, where the mean blood flow at rest is around 0.65 ml/min/cc, we do not expect flow to be above 1 at rest, so the SCF results with higher bounds would be difficult to interpret, since the higher values could also be due to spill-over from blood-pools, or
voxels actually containing blood or noise.}

{In terms of computation, i}t took about 4 hours to complete all 6000 iterations for each noise realization {of simulated data}, {using Matlab R2014b, running on a single node of the Linux computational cluster Katana at UNSW, Australia. This is equivalent to running on an average PC. The total number of voxels was 5746}. We found that all the parameters converged quite quickly.  {For SCF and SKMS, the computational time was around 1 minute.
For the pig study data involving 16821 voxels, and a longer time series involving 29 time points, the computational times were 23.5 hours and 6 hours for SMM and SCF respectively.}
We note that although SMM is computationally more expensive, it provides additional uncertainty estimation, which the other two faster methods do not.  Parallel computation or other computational methods, such as variational Bayes \cite{attias2000variational}, can be adopted to further speed up this process.

In the future, there are three directions to be considered to further develop our approach. First, we can relax the within cluster homogeneity assumption. This is easily achievable by relaxing the mean of the normal mixture to allow them to vary for each voxel observation. However, this substantially increases the number of parameters that needs to be estimated and presents a computational challenge.
Second, we can consider the use of sinogram data rather than reconstructed data to estimate kinetic parameters, { and this can reduce the additional noise introduced through the reconstruction step; however, this approach can be computationally challenging for full Bayesian analyses}. Third, given its flexibility, SMM can be easily extended to more advanced kinetic models, such as the two-compartment tissue model without too much modification. 
{Although we assess performance using simulations of cardiac perfusion PET imaging and demonstrate in vivo data for this application, }our approach is not limited to this specific context and may also benefit other dynamic PET procedures as well as dynamic SPECT, dynamic contrast enhanced CT (DCE-CT), and dynamic contrast enhanced MR (DCE-MR).

\section{Conclusion}\label{sec:conclusion}

This paper {proposed} a novel spatiotemporal approach, SMM, to infer parametric PET images. By borrowing information from nearby voxels, SMM can be used to simultaneously estimate kinetic parameters and classify voxels with similar kinetic parameters into spatially homogeneous groups. We adopted the MRF to incorporate the spatial dependence of voxels. We developed an efficient MCMC algorithm for the computation, which estimates
all unknown parameters, including the notoriously difficult spatial smoothness parameter $\beta$ in the Potts model. The method provides  parameter uncertainty estimation, as well as a principled way
to determine the optimal number of voxel groups.
We used simulated {cardiac} perfusion PET data to evaluate the performance of SMM and compared them with SCF and SKMS. SMM was substantially less sensitive to noise than the other methods, it also yielded an overall smaller bias than SCF and SKMS.  
In the pig study data,  SMM was shown to produce smoother parametric images compared to the standard curve fitting. Although simulation and experimental data were based on cardiac PET studies of a one-compartment model, the approach may benefit other dynamic PET procedures, as well as more complex compartmental models.

\section*{Acknowledgements}
This research was supported in part by UNSW 2014 Science Silver Star grant and NIH grants R01-HL118261, R01-HL110241, and NIH T32EB013180.

\newpage

\subsection*{Appendix}

\subsubsection*{Markov chain Monte Carlo}
We use MCMC for sampling from {the joint posterior distribution of ${\boldsymbol z}$ and all other parameters},  given by Equation \ref{eqn:post}.
The prior distribution $f({\boldsymbol \mu}, {\boldsymbol \Sigma}, \beta)$ is taken as product of the individual prior components 
$f(K^1_{1}), \ldots,  f(K^{G-1}_{1}), f(k^{1}_{2}), \ldots, f(k^{G-1}_{2}),
f(\mu_{g^*}^1),\ldots,  f(\mu_{g^*}^T),\\
 f(\sigma^{2,1}),\ldots, f(\sigma^{2,T}), f(\beta)$,
as defined in Section \ref{sec:priors}.

The first term on the right side of  Equation \ref{eqn:post} is given by
 \begin{align*}
&f(  y_i  | z_i=g, {\boldsymbol \mu}_g, {\boldsymbol \Sigma} ) \\
& = (2\pi)^{-T/2}|{\boldsymbol \Sigma} |^{-1/2} \exp(-\frac{1}{2} (y_i - {\boldsymbol \mu}_g)' {\boldsymbol \Sigma}^{-1}(y_i - {\boldsymbol \mu}_g) ),
\end{align*}
and the second term on the right side of the equation is given by
\begin{equation*}
  f({\bf z}|\beta)=\frac{1}{C(\beta)}\exp\{\beta  \sum_{i \sim j} I(z_i = z_j) \}.
\end{equation*}
This is the Potts model, where ${\bf z}=(z_1,\ldots, z_n)$, $I(\cdot)$ denotes indicator function taking value 1 if $z^{(l-1)}_j =g$ and 0 otherwise, {and $i\sim j$ denotes the voxels $j$ in the neighbourhood of voxel $i$}. The partition function $C(\beta)$ is estimated offline using thermal dynamic integration. \cite{green2002hidden} {We use an 8 nearest neighbour first order structure for the Potts model.}\\

Our computational algorithm {for the one-compartmental model in Equation \ref{eqn:kinetic}} proceeds as follows: 
\begin{description}
\item {\bf Step 1} Set $l=1$ and initialise parameters $K^{1,(0)}_{1}, k^{1, (0)}_{2},\ldots, K^{G-1, (0)}_{1}, k^{G-1, (0)}_{2},  {\boldsymbol \mu}^{(0)}_{g^*}, \sigma^{2, 1,(0)}, \\ \ldots, \sigma^{2, T,(0)}, {\bf z}^{(0)}, \beta^{(0)}$.
\item {\bf Step 2} Update $K^g_{1}$, for $g=1,\ldots, G-1$. Simulate a new value
$$ K^{'g}_{1}  \sim N(  K^{g, (l-1)}_{1}, \delta_{K_1}^2)$$
and compute ${\boldsymbol \mu}'_g$ with $K^{' g}_{1}$, according to Equation (\ref{eqn:kinetic}). Set $K^{g, (l)}_{1}$ to $ K^{' g}_{1} $ with probability $\alpha$, where
\begin{align*}
 & \alpha= \min  \{ 1, \alpha^{\star} \}
\end{align*}
 with $$\alpha^{\star}=  \frac{\prod_{i\in \{i: z^{(l-1)}_i=g\} } f(y_i|z^{(l-1)}_i, {\boldsymbol \mu}'_g, {\boldsymbol \Sigma}^{(l-1)} )  f(K^{'g}_{1})}{\prod_{i\in \{i: z^{(l-1)}_i=g\} } f(y_i|z^{(l-1)}_i, {\boldsymbol \mu}^{(l-1)}_g, {\boldsymbol \Sigma}^{(l-1)} )f(K_{1}^{g, (l-1)})}$$.

Otherwise, set $K^{g, (l)}_{1}$ to $ K^{g, (l-1)}_{1} $.

\item {\bf Step 3} Update $k^g_{2}$,  For $g=1,\ldots, G-1$. Analogously to Step 2.

\item {\bf Step 4} Update $\mu_{g^*}^t$, for $t=1,\ldots, T$. Simulate a new value
$$  \mu_{g^*}^{'t} \sim N(  \mu_{g^*}^{t, (l-1)}, \delta_{\mu_{g^*}}^2).$$
Set $\mu^{t, (l)}_{g^*}$ to $  \mu_{g^*}^{'t} $ with probability $\alpha$, where
$$\alpha= \min \{ 1, \alpha^{\star} \}$$
with
$$\alpha=  \frac{\prod_{i\in \{i: z^{(l-1)}_i=g^*\} } f(y_i|z^{(l-1)}_i, {\boldsymbol \mu}'_{g^*}, {\boldsymbol \Sigma}^{(l-1)} )  f(\mu_{g^*}^{'t}) }{\prod_{i\in \{i: z^{(l-1)}_i=g\} } f(y_i|z^{(l-1)}_i, {\boldsymbol \mu}^{(l-1)}_{g^*}, {\boldsymbol \Sigma}^{(l-1)} )f(\mu_{g^*}^{t, (l-1)}))} . $$
Otherwise, set $\mu_{g^*}^{t, (l)}$ to $\mu_{g^*}^{t, (l-1)}$.

\item {\bf Step 5}   Update $\sigma^{2,t}$, for $t=1,\ldots, T$.  Simulate from the Inverse Gamma distribution
$$\sigma^{2, t, (l)} \sim  IG \left(n/2+a,  \frac{1}{2}\sum_{i=1}^n(y_{i}^t- {\boldsymbol \mu}_g^{t, (l)})^2 +b\right).$$

\item {\bf Step 6} Update ${\bf z}$.  Each $i=1,\ldots, N$, compute
\begin{align*}
 & w_g = MVN(y_i;f(K_{1}^{g},k_{2}^{g}), {\boldsymbol \Sigma})  \exp\{\beta^{(l-1)} \sum_{j,j\in \partial i}I(z^{(l-1)}_j =g)\}, \quad g=1,\ldots G,
\end{align*}

and normalise $w'_g = w_g/\sum_{g=1}^G w_g$, where $f(K_{1}^{g},k_{2}^{g})$ denotes Equation \ref{eqn:kinetic}. $\partial i$ denotes the set of neighbours of vertex $i$.  Set $z_i^{(l)}$ according to the Multinomial distribution
$$z_i^{(l)} \sim MN(w'_1,\ldots, w'_G).$$

\item {\bf Step 7} Update $\beta$. Simulate a new value
$$ \beta'  \sim N(\beta^{(l-1)}, \delta_{\beta}^2)$$ and set $\beta^{(l)}$ to $\beta'$ with probability $\alpha$, where
$$\alpha= \mbox{ min } \left\{ 1, \frac{ f({\bf z}^{(l)} | \beta') f(\beta')}{ f({\bf z}^{(l)} | \beta^{(l-1)})f(\beta^{(l-1)})}  \right\}.$$
Otherwise, set $\beta^{(l)}$ to $ \beta^{(l-1)}$.

\item {\bf Step 8} set $l=l+1$, if $l < L$, go to Step 2.

\end{description}

  \newpage
%\input{version-medphys.bbl}
%
% \bibliographystyle{chicago}
% % \setcitestyle{numbers}
%  \bibliography{ref}
% \bibliography{aipsamp}
  % \bibliographystyle{biblatex}

\end{document}